\documentclass{article}

\usepackage{arxiv}

\usepackage[utf8]{inputenc} 
\usepackage[T1]{fontenc}    
\usepackage{hyperref}       
\usepackage{url}            
\usepackage{booktabs}       
\usepackage{amsfonts}       
\usepackage{nicefrac}       
\usepackage{microtype}      
\usepackage{lipsum}		
\usepackage{graphicx}
\usepackage{natbib}
\usepackage{doi}

\usepackage{lscape}
\newcommand{\blandscape}{\begin{landscape}}
\newcommand{\elandscape}{\end{landscape}}
\usepackage{float}
\usepackage{longtable}
\usepackage{array}
\usepackage{multirow}
\usepackage{wrapfig}
\usepackage{colortbl}
\usepackage{pdflscape}
\usepackage{tabu}
\usepackage{threeparttable}
\usepackage{threeparttablex}
\usepackage[normalem]{ulem}
\usepackage{makecell}
\usepackage{xcolor}
\usepackage{bookmark}
\usepackage{amsmath} 

\title{Improving Power in Randomized Controlled Trials with Time-to-Event Endpoints: A Risk-Free Approach}
\author{
    Junyi Zhou \\
	Amgen Inc.\\
	One Amgen Center Drive\\
	Thousand Oaks, CA, 91320 \\
	\texttt{jzhou10@amgen.com} \\
\And 
    Qing Liu \\
	Amgen Inc.\\
	One Amgen Center Drive\\
	Thousand Oaks, CA, 91320 \\
	\texttt{qliu02@amgen.com} \\
\And 
    May Mo \\
	Amgen Inc.\\
	One Amgen Center Drive\\
	Thousand Oaks, CA, 91320 \\
	\texttt{mmo@amgen.com} \\
\And 
    Amy Xia \\
	Amgen Inc.\\
	One Amgen Center Drive\\
	Thousand Oaks, CA, 91320 \\
	\texttt{hxia@amgen.com} \\
}
\date{}

\begin{document}
\maketitle

\begin{abstract}
Leveraging external or historical data to improve the efficiency of randomized clinical trials without introducing bias or inflating the Type I error rate remains challenging. Recent work on externally trained prognostic scores, such as PROCOVA for continuous endpoint, has demonstrated a risk-free approach via covariate adjustment. However, extending this paradigm to time-to-event endpoints is nontrivial due to the non-collapsibility of the unconditional hazard ratio (HR). In this paper, we address this challenge by proposing a unified framework for incorporating complex, high-dimensional prognostic information learned from external data into the primary analysis of RCTs with time-to-event endpoints, while targeting the unconditional hazard ratio. The proposed procedure proceeds in two steps. First, a prognostic score is estimated from external or historical data by regressing martingale residuals on baseline covariates using flexible supervised learning methods. Second, the fitted score is included as an additional covariate in the nonparametric covariate-adjusted log-rank test and the associated unconditional HR estimator of \citet{ye2024covariate}. The proposed method controls Type I error and provides asymptotic unbiased estimation of the unconditional HR, irrespective of prognostic model misspecification or population heterogeneity between external/historical and trial data. We show that the variance reduction, and corresponding event count savings, are approximately equal to the squared correlation between the prognostic score and the martingale pseudo-outcome in the trial. Simulation studies demonstrate satisfactory finite-sample performance and meaningful efficiency gains when historical prognostic information is informative.
\end{abstract}

\keywords{covariate adjustment \and hazard ratio \and external information borrowing \and historical data \and prognostic score \and Type~I error control \and randomized trials}

\section{Introduction}\label{introduction}
 
Randomized controlled trials (RCTs) remain the gold standard for
establishing the causal efficacy of new treatments, yet they are
simultaneously among the most resource-intensive components of drug
development. The dual burden of high per-patient costs and increasingly
challenging enrollment---particularly acute in rare diseases, pediatric
populations, and oncology where eligible patients populations are
small or randomization to control arm may raise ethical concerns. 
Such constraints have motivated sustained methodological interest 
in approaches that improve statistical efficiency without requiring 
additional participants\citep{bentley2019conducting}. At its core, 
the challenge is one of precision: for a fixed sample size, any reduction 
in the variance of the treatment effect estimator translates directly into 
increased power or, equivalently, a reduced sample size requirement for 
a fixed power target. Incorporating information from existing data sources, 
such as historical trial data, real-world data (RWD), or external data 
from observational studies, offers a natural avenue for precision improvement.
 
The dominant paradigm for incorporating historical data into RCT
is Bayesian dynamic borrowing (BDB), which links historical
data with RCT data through an informative prior \citep{Viele2014-lj}. 
The amount of information borrowed is adaptively adjusted through parameters 
(e.g., discounting factors or shrinkage parameters) that depend on the degree 
of homogeneity between external/historical data and RCT data. Various instantiations 
of this idea have been proposed, including mixture prior \citep{schmidli2014robust},
commensurate priors \citep{hobbs2011hierarchical}, and Bayesian hierarchical models 
\citep{han2017covariate, lewis2019borrowing}. More recently, propensity-score-weighted 
BDB methods have been proposed to account for covariate distribution differences between 
historical and concurrent populations \citep{chen2023leveraging, wang2024propensity}.
 
Despite their appeal, BDB approaches share fundamental limitations:
uncontrolled inflation of the Type~I error and bias in treatment effect estimates. 
The mechanism underlying this inflation is structural: BDB methods incorporate 
historical outcomes directly into the likelihood or prior for the treatment effect 
parameter, so the null distribution of the test statistic is sensitive to whether 
the historical population is exchangeable with the current trial---a condition 
that cannot be verified prospectively. 
\citet{kopp2020power} proved under very general conditions that no borrowing 
procedure can simultaneously achieve power gains and maintain the Type~I error at its nominal level. 

While the FDA's 2026 guidance on the use of Bayesian methodology in clinical trials acknowledges that, 
in limited settings with strong scientific justification and demonstrated relevance and comparability 
between external/historical and current date, formal Type-I error calibration may not be required, 
it nevertheless emphasizes extensive pre-trial evaluations of design operating characteristics. 
These include assessment of expected bias and MSE, Bayesian power,  probabilities of erroneous decision-making, expected interval width, 
and comprehensive sensitivity analyses to prior-data conflict. Together, these requirements 
substantially limit the practical use of BDB in confirmatory settings broadly.

A fundamentally different strategy for leveraging external or historical data is to use such data to construct a "super covariate" \citep{holzhauer2023super} that improves RCT efficiency via covariate adjustment. The externally trained super covariate would outperform conventional adjustment using individual covariates by capturing complex prognostic information using machine learning models from rich external data. By integrating a large number of baseline covariates into a single adjustment, this approach also achieves effective dimension reduction, thereby improving feasibility in RCTs with limited sample size. 

Prognostic covariate adjustment (PROCOVA) is one representative approach in this type of strategy\citep{schuler2022increasing}. The idea is to use historical
patient-level data, to train a prognostic model---a predictor of the control-group outcome as a function of baseline covariates. This pre-trained prognostic score is then included as an additional baseline covariate in the primary
analysis of the RCT. The theoretical foundation of PROCOVA rests on the robustness properties of the analysis of covariance (ANCOVA) model for continuous endpoints in RCTs. The ANCOVA-based treatment
effect estimator is unbiased for the unconditional treatment effect
regardless of whether the prognostic model is correctly specified,
provided that randomization is employed
\citep{tsiatis2008covariate, zhang2008improving, lin2013agnostic, ye2023toward}. Notably, the mechanism by which the historical
information enters the analysis is entirely different from BDB: the
external information is encoded only as a baseline covariate, not as
part of the likelihood or prior for the treatment effect.
Consequently, Type~I error control is guaranteed by randomization,
regardless of how different the historical population is from the trial
population and regardless of whether the prognostic model is correctly
specified \citep{liao2025prognostic, hojbjerre2025tutorial}. Any
predictive power in the score reduces residual outcome variability and
thereby increases precision---a ``free lunch'' in the sense that the
worst case is simply no gain. The PROCOVA methodology has received a
qualification opinion from the European Medicines Agency
\citep{EMA_qualification}.
More broadly, the U.S. Food and Drug Administration's 2023 guidance on 
covariate adjustment in RCTs \citep{FDA2023_CovAdj} endorses baseline 
covariate adjustment as a means to improve efficiency without inflating Type~I error.
A series of simulation and empirical studies,
including \citet{walsh2021using} and \citet{burman2024digital}, have
confirmed the practical utility of the approach. \citet{hojbjerre2025tutorial} also provided a detailed tutorial with practical guidance. 

Several extensions of PROCOVA have been developed in recent years. \citet{liao2025prognostic} generalized the framework to semiparametric efficient estimators and targeted maximum likelihood estimation (TMLE). While more recently, \citet{hojbjerre2026powering} extended prognostic score adjustment to generalized linear models. Despite these important advances, extending the framework to time-to-event endpoints remains particularly challenging. In settings where the hazard ratio (HR) is the primary estimand, fundamental issues arise that are not addressed by existing approaches. The most natural analogue---including a prognostic score as an additional covariate in a Cox proportional hazards model---fails to preserve the risk-free property. The HR is a non-collapsible effect
measure \citep{agresti2013categorical, daniel2021making}: when
additional covariates are added to a Cox model, the conditional HR
converges to a quantity that is numerically different from the unconditional
HR, even when the model is correctly specified or randomization is employed
\citep{karrison2018restricted, siegfried2023relevance}. As a result,
na\"ive covariate adjustment in a Cox model changes the target estimand 
from unconditional HR to the conditional HR, 
a point explicitly acknowledged in the FDA's 2023 covariate adjustment guidance, 
which cautions that the hazard ratio can be non-collapsible and that sponsors should 
prespecify whether the estimand of interest is a conditional or unconditional treatment effect \citep{FDA2023_CovAdj}. 
Indeed, \citet{siegfried2023relevance} explicitly noted that extending
prognostic score adjustment to proportional hazards models is ``not
straightforward'' due to non-collapsibility. 
 
The purpose of this paper is to resolve this gap by proposing a
risk-free framework for borrowing external/historical information in RCTs with time-to-event endpoints and the unconditional HR as
the primary estimand via prognostic score adjustment. Our approach proceeds in two steps. First, we
adopt the covariate-adjusted log-rank test and its companion
covariate-adjusted HR estimator, recently developed by
\citet{ye2024covariate}, as the analysis framework. This framework inherits 
the key robustness properties that ANCOVA
provides for continuous endpoints: the covariate-adjusted log-rank
test is valid and the HR estimator is consistent for the unconditional HR
estimand regardless of whether any covariate model is correctly
specified, and both are asymptotically at least as efficient as
their unadjusted counterparts under any covariate-adaptive
randomization scheme satisfying mild regularity conditions. Second, we
propose to include a prognostic score pre-trained on external/historical
control data as an additional baseline covariate, directly analogous to
the PROCOVA approach.
Because the framework of \citet{ye2024covariate}
is entirely model-free---it requires neither correct specification of
the hazard function nor of the censoring distribution---adding the
prognostic score cannot inflate the Type~I error, regardless of the
quality of the prognostic model. At the same time, any genuine
predictive signal in the score reduces the variability of the
covariate-adjusted log-rank score function and thereby increases both
the power of the test and the precision of the HR estimator.

 To the best of our knowledge, no existing work has proposed or
rigorously validated the integration of a prognostic score trained on external/historical data
to increase the precision of HR for time-to-event endpoints. 
The closest related works is \citet{hattori2022sample}, who developed a sample size formula for
augmented log-rank tests by leveraging the correlation between baseline
covariates and martingale residuals. However, they do not use this external
information to increase the precision of unconditional HR estimation.
Also, their approach adjusts directly for observed covariates
rather than a prognostic score pre-trained on external data using
flexible learning methods, and does not address the dimension reduction
and non-linear prognostic modelling that becomes valuable when
historical datasets are rich and high-dimensional.
Moreover, they do not address Type~I error
control when the historical population is not exchangeable with the
trial population. Our work provides this
resolution through the nonparametric covariate-adjusted framework of
\citet{ye2024covariate}, which maintains the unconditional HR as the
estimand and preserves full Type~I error control.

The remainder of this paper is organized as follows. Section~2
introduces the notation, reviews the covariate-adjusted log-rank test
of \citet{ye2024covariate}, the covariate-adjusted HR estimator,
and its asymptotic properties. Section~3 presents the
external information borrowing mechanism through a prognostic score,
derives the variance reduction and event savings formulas, and discusses
the training of the prognostic model from historical data. Section~4
presents simulation studies evaluating finite-sample performance across
a range of settings that vary in the quality and relevance of the
external data. 
Section~5 illustrates the method for increasing precision for a Phase~III colorectal cancer trial
using data from the control arm of a historical study.
Section~6 concludes with a discussion of practical
implications, extensions, and directions for future work.

\section{Covariate-Adjusted Hazard Ratio Estimator}\label{covariate-adjusted-hazard-ratio-estimator}
 
In this section we introduce a robust framework for covariate adjustment
in the analysis of time-to-event outcomes with the unconditional HR as the
target estimand. To be self-contained, we begin by reviewing the covariate-adjusted log-rank
test of \citet{ye2024covariate}, which provides the foundation, and then
naturally dive into the corresponding covariate-adjusted HR estimator and 
its asymptotic properties. Notation follows \citet{ye2024covariate}
throughout.
 
\subsection{Notation and Setup}\label{notation-and-setup}
 
Consider a two-arm RCT with $n$ patients. For
patient $i = 1, \ldots, n$, let $I_i \in \{0, 1\}$ denote the
treatment indicator ($I_i = 1$ for the experimental treatment,
$I_i = 0$ for control), and let $X_i \in \mathbb{R}^p$ denote a vector
of observed baseline covariates. Let $T_{ij}$ and $C_{ij}$ denote,
respectively, the potential failure time and potential censoring time
for patient $i$ under treatment $j \in \{0,1\}$. The observed data for
patient $i$ are $\{\widetilde{T}_i, \Delta_i\}$, where
$\widetilde{T}_i = I_i \min(T_{i1}, C_{i1}) + (1-I_i)\min(T_{i0},C_{i0})$
is the observed event or censoring time and
$\Delta_i = I_i \mathbf{1}(T_{i1} \le C_{i1}) + (1-I_i)\mathbf{1}(T_{i0} \le C_{i0})$
is the event indicator. Let $\tau > 0$ be a prespecified analysis
horizon satisfying $\Pr\{\min(T_{ij}, C_{ij}) \ge \tau\} > 0$ for
$j = 0, 1$.
 
Define the counting process
$N_{ij}(t) = \mathbf{1}(\widetilde{T}_i \le t,\, \Delta_i = 1,\, I_i = j)$
and the at-risk indicator
$Y_{ij}(t) = \mathbf{1}(\widetilde{T}_i \ge t,\, I_i = j)$ for
treatment arm $j$, with $N_i(t) = N_{i1}(t) + N_{i0}(t)$ and
$Y_i(t) = Y_{i1}(t) + Y_{i0}(t)$. Sample-average processes are denoted
\[
\bar{Y}_j(t) = n^{-1}\sum_{i=1}^n Y_{ij}(t), \quad
\bar{Y}(t) = \bar{Y}_1(t) + \bar{Y}_0(t), \quad
\bar{N}(t) = n^{-1}\sum_{i=1}^n N_i(t).
\]
 
Let $\pi \in (0,1)$ be the prespecified allocation proportion to the
experimental treatment, so that $\Pr(I_i = 1) = \pi$ for all $i$. We
adopt the following standard conditions.
 
\textbf{Assumption 1} (Non-informative censoring).
$C_{ij} \perp T_{ij} \mid I_i$ for $j = 0, 1$.

\textbf{Condition 1} (Simple randomization).
The treatment indicators $I_1, \ldots, I_n$ are i.i.d.\ with
$\Pr(I_i = 1) = \pi$, and are jointly independent of all potential
outcomes $\{(T_{i0}, T_{i1}, C_{i0}, C_{i1})\}_{i=1}^n$.

Sections~2 and~3 are developed under simple randomization throughout.
The extension to stratified and covariate-adaptive randomization,
which follows in close parallel with the results of Sections~2 and~3,
is presented in Appendix~\ref{variance-reduction-and-event-savings-under-stratified-randomization}.
 
\textbf{Target estimand.} We target the unconditional log-HR
$\theta_0$, defined as the unique solution to the population-level Cox
score equation \[
E\!\left[\int_0^\tau \left\{I_i - \frac{e^{\theta_0}\bar{Y}_1(t)}{e^{\theta_0}\bar{Y}_1(t) + \bar{Y}_0(t)}\right\}dN_i(t)\right] = 0
\] in a model with treatment as the sole covariate. Under the
proportional hazards assumption $\lambda_1(t) = e^{\theta_0}\lambda_0(t)$,
the quantity $e^{\theta_0}$ equals the common HR; without
proportional hazards, $e^{\theta_0}$ remains a
weighted average of time-varying HRs \citep{lin1989robust, struthers1986misspecified}. We write
$\hat{\theta}_L$ for the standard maximum partial likelihood estimator
(with treatment as the only covariate), which converges in probability
to $\theta_0$.

We emphasize that $\theta_0$ is a \textbf{unconditional} log-HR:
it is defined without conditioning on any baseline covariate. This
contrasts with the conditional log-HR arising from a Cox model that
includes additional covariates, which targets a numerically different
quantity due to the non-collapsibility of the HR \citep{daniel2021making}.
Maintaining the unconditional HR as the estimand is essential for the
risk-free property developed in Section~3.

\subsection{The Covariate-Adjusted Log-Rank Test}
\label{the-covariate-adjusted-log-rank-test}
 
We briefly review the covariate-adjusted log-rank test of
\citet{ye2024covariate}, which provides the analytical foundation for
Section~\ref{the-covariate-adjusted-hr-estimator}. The standard
(unadjusted) log-rank test statistic is $T_L = n^{1/2}\hat{U}_L /
\hat{\sigma}_L$, where
\[
  \hat{U}_L
  = \frac{1}{n}\sum_{i=1}^n \int_0^\tau
      \!\left(I_i - \frac{\bar{Y}_1(t)}{\bar{Y}(t)}\right)dN_i(t),
  \qquad
  \hat{\sigma}^2_L
  = \frac{1}{n}\sum_{i=1}^n \int_0^\tau
      \frac{\bar{Y}_1(t)\bar{Y}_0(t)}{\bar{Y}(t)^2}\,dN_i(t).
\]
 
The key step is to linearize $\hat{U}_L$ via a Taylor expansion:
\[
  \hat{U}_L
  = \frac{1}{n}\sum_{i=1}^n \bigl\{I_i O_{i1} - (1-I_i)O_{i0}\bigr\}
    + o_p(n^{-1/2}),
\]
where the pseudo-outcomes $O_{ij}$ ($j = 0, 1$) are defined
as
\begin{equation}\label{eq:pseudo_pop}
  O_{ij}
  = \int_0^\tau
      \{1 - \mu(t)\}^j\,\{\mu(t)\}^{1-j}
      \bigl\{dN_{ij}(t) - Y_{ij}(t)\,p(t)\,dt\bigr\},
\end{equation}
with $\mu(t) = \mathbb{E}\{I_i \mid Y_i(t)=1\}$ the treatment
proportion among patients at risk at time $t$, and
$p(t)\,dt = \mathbb{E}\{dN_i(t)\}/\mathbb{E}\{Y_i(t)\}$ the marginal
event rate among those at risk. Sample analogues of $O_{ij}$ are
\begin{equation}\label{eq:PseudoO}
  \hat{O}_{ij}
  = \int_0^\tau
      \frac{\bar{Y}_{1-j}(t)}{\bar{Y}(t)}
      \left[dN_{ij}(t)
            - Y_{ij}(t)\frac{d\bar{N}(t)}{\bar{Y}(t)}\right],
  \quad j = 0, 1.
\end{equation}
 
Because $\hat{U}_L$ is now expressed as an average of i.i.d.\ terms,
standard regression augmentation applies
\citep{tsiatis2008covariate}: subtracting the linear projection of
$O_{ij}$ onto $X_i$ reduces variance without introducing bias.
Specifically, define arm-specific regression coefficients
\[
  \hat{\beta}_j
  = \Bigl(\sum_{i:\,I_i=j}
      (X_i-\bar{X}_j)(X_i-\bar{X}_j)^\top\Bigr)^{-1}
    \sum_{i:\,I_i=j}(X_i-\bar{X}_j)\hat{O}_{ij},
  \quad j = 0, 1,
\]
where $\bar{X}_j$ is the sample mean of $X_i$ in arm $j$. The
covariate-adjusted log-rank score is
\[
  \hat{U}_{CL}
  = \hat{U}_L
    - \frac{1}{n}\sum_{i=1}^n
        \bigl\{I_i(X_i-\bar{X})^\top\hat{\beta}_1
               - (1-I_i)(X_i-\bar{X})^\top\hat{\beta}_0\bigr\},
\]
and the corresponding covariate-adjusted variance estimator is
\[
  \hat{\sigma}^2_{CL}
  = \hat{\sigma}^2_L
    - \hat{\pi}(1-\hat{\pi})
      (\hat{\beta}_1+\hat{\beta}_0)^\top
      \hat{\Sigma}_X
      (\hat{\beta}_1+\hat{\beta}_0),
\]
where $\hat{\pi} = n_1/n$ and $\hat{\Sigma}_X$ is the sample
covariance matrix of $X_i$. The covariate-adjusted log-rank test
statistic is $T_{CL} = n^{1/2}\hat{U}_{CL}/\hat{\sigma}_{CL}$.
 
Since the variance reduction term is positive semidefinite,
$\operatorname{var}(\hat{U}_{CL}) \le \operatorname{var}(\hat{U}_L)$,
with strict inequality whenever $X_i$ has non-zero linear association
with the pseudo-outcomes. Crucially, this efficiency guarantee holds
without any model assumption on the failure or censoring distributions
beyond Assumption~1 and Condition~1: the validity of $T_{CL}$ is
secured entirely by randomization. It is precisely this model-free
robustness that enables risk-free information borrowing in
Section~\ref{risk-free-information-borrowing-via-prognostic-score}.

\subsection{The Covariate-Adjusted HR Estimator}
\label{the-covariate-adjusted-hr-estimator}
 
We now derive a covariate-adjusted estimator of the unconditional 
log-HR $\theta_0$, exploiting the natural connection between the
log-rank score function and the Cox partial likelihood score. This
connection was established in \citet{ye2024covariate} (Section~3 and
Supplementary Material, Section~S2.2). Here we provide a self-contained
derivation and present the main results, as the estimator and its asymptotic
properties are central to the prognostic score adjustment framework of
Section~\ref{risk-free-information-borrowing-via-prognostic-score}.

\subsubsection{From the Adjusted Score to the Adjusted Estimator}
\label{from-the-adjusted-score-to-the-adjusted-estimator}
 
For a general parameter value $\vartheta$, the unadjusted Cox partial
likelihood score with treatment as the sole covariate is \[
\hat{U}_L(\vartheta) = \frac{1}{n}\sum_{i=1}^n \int_0^\tau \left(I_i - \frac{e^\vartheta \bar{Y}_1(t)}{e^\vartheta \bar{Y}_1(t) + \bar{Y}_0(t)}\right)dN_i(t).
\] Note that $\hat{U}_L(0) = \hat{U}_L$ as defined in Section~2.2, and
the standard maximum partial likelihood estimator $\hat{\theta}_L$
solves $\hat{U}_L(\vartheta) = 0$.
 
The covariate adjustment proceeds by augmenting $\hat{U}_L(\vartheta)$
with an additive correction term that exploits the linear relationship
between $X_i$ and arm-specific pseudo-outcomes derived from the score at
a given $\vartheta$. Specifically, define \[
\hat{O}_{i1}(\vartheta) = \int_0^\tau \frac{\bar{Y}_0(t)}{e^\vartheta\bar{Y}_1(t) + \bar{Y}_0(t)}\left[dN_{i1}(t) - Y_{i1}(t)\frac{e^\vartheta\,d\bar{N}(t)}{e^\vartheta\bar{Y}_1(t) + \bar{Y}_0(t)}\right],
\] \[
\hat{O}_{i0}(\vartheta) = \int_0^\tau \frac{e^\vartheta\bar{Y}_1(t)}{e^\vartheta\bar{Y}_1(t) + \bar{Y}_0(t)}\left[dN_{i0}(t) - Y_{i0}(t)\frac{d\bar{N}(t)}{e^\vartheta\bar{Y}_1(t) + \bar{Y}_0(t)}\right].
\] These are the natural generalizations of $\hat{O}_{ij}$ (defined at
$\vartheta = 0$) to arbitrary $\vartheta$, and they satisfy the
algebraic identity
$\hat{U}_L(\vartheta) = n^{-1}\sum_i\{I_i\hat{O}_{i1}(\vartheta) - (1-I_i)\hat{O}_{i0}(\vartheta)\}$.
Arm-specific regression coefficients at $\vartheta$ are \[
\hat{\beta}_j(\vartheta) = \left(\sum_{i:\,I_i=j}(X_i-\bar{X}_j)(X_i-\bar{X}_j)^\top\right)^{-1}\sum_{i:\,I_i=j}(X_i-\bar{X}_j)\hat{O}_{ij}(\vartheta), \quad j=0,1.
\]
 
The covariate-adjusted score at $\vartheta$ is then
\begin{equation} \label{eq:CL_Score}
\hat{U}_{CL}(\vartheta) = \hat{U}_L(\vartheta) - \frac{1}{n}\sum_{i=1}^n\bigl\{I_i(X_i-\bar{X})^\top\hat{\beta}_1(\hat{\theta}_L) - (1-I_i)(X_i-\bar{X})^\top\hat{\beta}_0(\hat{\theta}_L)\bigr\}.
\end{equation}
 
The covariate-adjusted HR estimator $\hat{\theta}_{CL}$ is
defined as the solution to 
\begin{equation}
\hat{U}_{CL}(\vartheta) = 0. \label{eq:Score}
\end{equation}

\subsubsection{Asymptotic Properties}
\label{asymptotic-properties}
 
The asymptotic theory follows from standard M-estimation arguments
applied to the linearized score.
 
\textbf{Consistency.} Because
$\hat{U}_{CL}(\vartheta) = \hat{U}_L(\vartheta) + o_p(1)$ uniformly
in $\vartheta$ in a neighborhood of $\theta_0$, and
$\hat{\theta}_L \to \theta_0$ in probability, a standard argument gives
$\hat{\theta}_{CL} \to \theta_0$ in probability.
 
\textbf{Asymptotic normality.} Applying a Taylor expansion to
$\hat{U}_{CL}(\vartheta)$ around $\theta_0$: \[
0 = \hat{U}_{CL}(\hat{\theta}_{CL}) = \hat{U}_{CL}(\theta_0) + \frac{\partial \hat{U}_{CL}(\vartheta)}{\partial\vartheta}\bigg|_{\vartheta=\bar\vartheta}(\hat{\theta}_{CL} - \theta_0),
\] where $\bar\vartheta$ lies between $\hat{\theta}_{CL}$ and
$\theta_0$. Since
$\partial\hat{U}_{CL}/\partial\vartheta = \partial\hat{U}_L/\partial\vartheta$,
the uniform law of large numbers \citep{andersen1982cox} gives \[
-\frac{\partial \hat{U}_L(\vartheta)}{\partial\vartheta}\bigg|_{\vartheta=\bar\vartheta} \xrightarrow{p} \sigma^2_L(\theta_0),
\] where
$\sigma^2_L(\theta) = \int_0^\tau \left[\frac{s^{(1)}(\theta,t)}{s^{(0)}(\theta,t)} - \left(\frac{s^{(1)}(\theta,t)}{s^{(0)}(\theta,t)}\right)^2\right] s^{(0)}(\theta,t)\,dt$,
with $s^{(0)}(\theta,t) = \{e^\theta\mu(t) + 1 - \mu(t)\}\mathbb{E}\{Y_i(t)\}$
and $s^{(1)}(\theta,t) = e^\theta\mu(t)\mathbb{E}\{Y_i(t)\}$ the
population-level risk-process sums.
 
For the numerator, by Theorem~1 of \citet{ye2024covariate} applied at
$\theta_0$ (where $\mathbb{E}[O_{i1}(\theta_0)] = \mathbb{E}[O_{i0}(\theta_0)] = 0$
by the definition of $\theta_0$ as the true parameter): \[
n^{1/2}\hat{U}_{CL}(\theta_0) \xrightarrow{d} N(0, \sigma^2_{CL}(\theta_0)),
\] where
$\sigma^2_{CL}(\theta_0) = \sigma^2_L(\theta_0) - \pi(1-\pi)(\beta_1(\theta_0) + \beta_0(\theta_0))^\top\Sigma_X(\beta_1(\theta_0) + \beta_0(\theta_0))$,
with $\beta_j(\theta_0) = \Sigma_X^{-1}\operatorname{cov}(X_i, O_{ij}(\theta_0))$.
This result holds under Assumption~1 and Condition~1, regardless of the
randomization scheme. Combining by Slutsky's theorem:
\begin{equation} \label{eq:CL_asym}
n^{1/2}(\hat{\theta}_{CL} - \theta_0) \xrightarrow{d} N\!\left(0,\, \frac{\sigma^2_{CL}(\theta_0)}{\{\sigma^2_L(\theta_0)\}^2}\right).
\end{equation}

\textbf{Variance estimation.} A consistent estimator of the asymptotic
variance in~\eqref{eq:CL_asym} is 
\begin{equation} \label{eq:VarEq}
\frac{\hat{\sigma}^2_{CL}(\hat{\theta}_{CL})}{\{\hat{\sigma}^2_L(\hat{\theta}_{CL})\}^2},
\end{equation} 
where \[
\hat{\sigma}^2_L(\hat{\theta}_{CL}) = -\frac{\partial\hat{U}_L(\vartheta)}{\partial\vartheta}\bigg|_{\vartheta=\hat{\theta}_{CL}} = \frac{1}{n}\sum_{i=1}^n\int_0^\tau \frac{e^{\hat{\theta}_{CL}}\bar{Y}_1(t)\bar{Y}_0(t)}{\{e^{\hat{\theta}_{CL}}\bar{Y}_1(t)+\bar{Y}_0(t)\}^2}dN_i(t)
\] is the observed Fisher information from the unadjusted Cox partial
likelihood, and \begin{equation} \label{eq:CL_Var}
\hat{\sigma}^2_{CL}(\hat{\theta}_{CL}) = \hat{\sigma}^2_L(\hat{\theta}_{CL}) - \hat{\pi}(1-\hat{\pi})\bigl(\hat{\beta}_1(\hat{\theta}_L) + \hat{\beta}_0(\hat{\theta}_L)\bigr)^\top\hat{\Sigma}_X\bigl(\hat{\beta}_1(\hat{\theta}_L) + \hat{\beta}_0(\hat{\theta}_L)\bigr).
\end{equation}
Consistency of~\eqref{eq:VarEq} follows from $\hat{\beta}_j(\hat{\theta}_L) \to \beta_j(\theta_0)$ in probability
and $\hat{\Sigma}_X \to \Sigma_X$ in probability, by the continuous mapping theorem. An asymptotic $(1-\alpha)$-level confidence interval for $\theta_0$ is
$\hat{\theta}_{CL} \pm z_{\alpha/2}\cdot\widehat{\operatorname{Var}}(\hat{\theta}_{CL})^{1/2}$, and a confidence interval for the HR $e^{\theta_0}$ is
obtained by exponentiating the endpoints.
 
All results, that is, consistency, asymptotic normality, guaranteed
efficiency gain, and validity of the variance estimator, hold
without any model assumption on the joint distribution of
$(T_i,\,C_i,\,X_i)$ beyond Assumption~1 and Condition~1. Neither
correct specification of any hazard model nor any assumption on the
censoring distribution is required. It is this model-free character
that is the essential prerequisite for the risk-free information
borrowing developed in
Section~\ref{risk-free-information-borrowing-via-prognostic-score}.

\section{Risk-Free Information Borrowing via Prognostic Score}\label{risk-free-information-borrowing-via-prognostic-score}
 
The critical property that we have established in Section~2 is that covariate adjustment 
operates at the score-function level while the estimand $\theta_0$ 
is defined entirely by the unadjusted population score equation 
and is therefore unaffected by the adjustment. This 
separation between inferential target and  variance-reduction 
mechanism allows externally trained prognostic scores to improve precision 
without altering the unconditional HR estimand or inflating Type~I error, 
regardless of the score construction or how different the external population is from the trial.

In this section, we show how a pre-trained prognostic
score from historical/external data can serve as the adjustment
covariate within this framework to increase precision in a
risk-free manner. We derive explicit formulas for the resulting variance
reduction and event count savings, and address the practical question of
how to train the prognostic model from historical/external data.

\subsection{Variance Reduction via Prognostic Score Adjustment}\label{variance-reduction-via-prognostic-score-adjustment}
 
We first characterize the efficiency gain as a function of the
prognostic score's association with the survival pseudo-outcomes. From
equation~\eqref{eq:CL_asym}, the asymptotic variance of
$\hat\theta_{CL}$ is \[
\text{Var}(\hat\theta_{CL}) = \frac{\sigma_{CL}^2(\theta_0)}{\sigma_L^4(\theta_0)} = \text{Var}(\hat\theta_{L}) \cdot \frac{\sigma_{CL}^2(\theta_0)}{\sigma_L^2(\theta_0)},
\] so the relative efficiency of the adjusted over the unadjusted
estimator is determined entirely by the ratio $\sigma_{CL}^2(\theta_0)/\sigma_L^2(\theta_0)$.

By Theorem~S2 of \citet{ye2024covariate}, the two population variances
satisfy
\begin{align}
  \sigma_L^{2}(\theta_0)
    &= \pi\,\nu_1 + (1-\pi)\nu_0,  \label{eq:sigL} \\
  \sigma_{CL}^{2}(\theta_0)
    &= \sigma_L^{2}(\theta_0)
       - \pi(1-\pi)(\beta_1(\theta_0)+\beta_0(\theta_0))^\top\Sigma_X(\beta_1(\theta_0)+\beta_0(\theta_0)),
       \label{eq:sigCL}
\end{align}
where $\nu_j = \operatorname{var}(O_{ij}(\theta_0))$ is the
marginal variance of the arm-$j$ pseudo-outcomes,
$\beta_j(\theta_0) = \Sigma_X^{-1}\operatorname{cov}(X_i,O_{ij}(\theta_0))$
is the population regression coefficient of $O_{ij}(\theta_0)$ on $X_i$, and
$\Sigma_X = \operatorname{var}(X_i)$.

\textbf{Scalar prognostic score.}
Now suppose $\eta(X_i)\in\mathbb{R}$ is a scalar prognostic score used
as the sole adjustment covariate (previous $X_i$ is replaced by $\eta(X_i)$). 
The population regression coefficient reduces to the scalar
$\beta_j(\theta_0) = \operatorname{cov}(\eta(X_i),O_{ij}(\theta_0))/\operatorname{var}(\eta(X_i))$,
and the numerator of the reduction term becomes
\begin{equation}\label{eq:num_reduction}
  \pi(1-\pi)(\beta_1(\theta_0)+\beta_0(\theta_0))^2\operatorname{var}(\eta(X_i))
  = \pi(1-\pi)
    \!\left(\sqrt{\nu_1(\theta_0)}\,\rho_1(\theta_0)+\sqrt{\nu_0(\theta_0)}\,\rho_0(\theta_0)\right)^{\!2},
\end{equation}
where
$\rho_j(\theta_0) = \operatorname{Corr}(\eta(X_i),O_{ij}(\theta_0))$
is the correlation between the prognostic score and the
arm-$j$ pseudo-outcome.
Equation~\eqref{eq:num_reduction} follows because
$\beta_j(\theta_0)\sqrt{\operatorname{var}(\eta)}
 = \operatorname{cov}(\eta(X_i),O_{ij}(\theta_0))/\sqrt{\operatorname{var}(\eta)}
 = \rho_j(\theta_0)\sqrt{\nu_j(\theta_0)}$.
Substituting into~\eqref{eq:sigL} and~\eqref{eq:sigCL}:
\begin{equation}\label{eq:VRR_exact}
  \frac{\sigma_{CL}^{2}(\theta_0)}{\sigma_L^{2}(\theta_0)}
  = 1 -
    \frac{\pi(1-\pi)
      \!\left(
        \rho_1(\theta_0)\sqrt{\nu_1(\theta_0)}
       +\rho_0(\theta_0)\sqrt{\nu_0(\theta_0)}
      \right)^{\!2}}
    {\pi\,\nu_1(\theta_0)+(1-\pi)\nu_0(\theta_0)}.
\end{equation}

\textbf{Variance reduction under the null and local alternatives.}
Formula~\eqref{eq:VRR_exact} involves $\nu_j(\theta_0)$ and
$\rho_j(\theta_0)$, which depend on the unknown true parameter $\theta_0$.
We now derive a clean closed-form expression under $H_0:\theta_0=0$ and
show it extends to local alternatives $\theta_0=O(n^{-1/2})$, which is
the practically relevant regime for trial planning.
 
Under $H_0$, the pseudo-outcomes $O_{ij}(0)$ reduce to
the standard log-rank pseudo-outcomes $O_{ij}$. At $\theta_0=0$, the population weights
$w_j(t;0)=s^{(1-j)}(0,t)/s^{(0)}(0,t)$ in the pseudo-outcome integral
become
\[
  w_1(t;0)
  = \frac{(1-\mu(t))E\{Y_i(t)\}}{E\{Y_i(t)\}}
  = 1-\mu(t),
  \qquad
  w_0(t;0)
  = \mu(t),
\]
where $\mu(t)=\Pr(I_i=1\mid Y_i(t)=1)$ is the at-risk treatment
proportion at time $t$. Under Condition~1 and Assumption~1, 
the risk-set composition satisfies $\mu(t)\to\pi$ as
$n\to\infty$ for all $t\in[0,\tau]$, so in large samples:
\[
  O_{i1}(0) = \int_0^\tau w_1(t;0)\,dM_{i1}(t)
  \;\;\to\;\;
  c_1(0)\,M_{i1}(\tau),
  \qquad
  O_{i0}(0) = \int_0^\tau w_0(t;0)\,dM_{i0}(t)
  \;\;\to\;\;
  c_0(0)M_{i0}(\tau),
\]
where at an arbitrary $\theta_0$, the population weight $w_j(t;\theta_0)$
 converges to the time-invariant scalar $c_j(\theta_0) = e^{\theta_0(1-j)}(1-\pi)^j\pi^{1-j} / (e^{\theta_0}\pi+(1-\pi))$, giving $c_1(0)=1-\pi$ and $c_0(0)=\pi$ at $\theta_0=0$.
$M_{ij}(t) = N_{ij}(t) - \int_0^t Y_{ij}(s)\, d\hat\Lambda(s)$ is the arm-j counting process martingale residual, evaluated under the pooled Nelson–Aalen estimator $\hat\Lambda$.
This martingale representation will play a key role in Section 3.3, where it motivates the
feasible choice of training target for the prognostic model. Consequently:
\begin{equation}\label{eq:nu_H0}
  \nu_1(0) = \operatorname{var}(O_{i1}(0)) \to (1-\pi)^2\,\nu_M,
  \qquad
  \nu_0(0) = \operatorname{var}(O_{i0}(0)) \to \pi^2\,\nu_M,
\end{equation}
where $\nu_M = \operatorname{var}(M_{ij}(\tau))$ is the common
within-arm martingale variance (equal across arms under $H_0$ by
symmetry). 

The corresponding arm-specific correlations satisfy
\begin{equation}\label{eq:rho_H0}
  \rho_j(0)
  = \operatorname{Corr}\!\bigl(\eta(X_i),\,O_{ij}(0)\bigr)
  = \operatorname{Corr}\!\bigl(\eta(X_i),\,c_j(0)\,M_{ij}(\tau)\bigr)
  = \operatorname{Corr}\!\bigl(\eta(X_i),\,M_{ij}(\tau)\bigr)
  =: \rho,
\end{equation}
and we use the fact that
correlations are invariant to positive scaling. The common value $\rho$
is the same for both arms under $H_0$ because under randomization
$X_i\perp I_i$, so $\operatorname{Corr}(\eta(X_i),M_{ij}(\tau))$
does not depend on $j$.
 
Substituting \eqref{eq:nu_H0} and \eqref{eq:rho_H0} into
\eqref{eq:VRR_exact}, we have
\begin{equation}\label{eq:VRR}
  \frac{\sigma_{CL}^{2}(\theta_0)}{\sigma_L^{2}(\theta_0)}
  = 1 - \frac{\pi(1-\pi)\,\rho^2\,\bigl(
    (1-\pi)\sqrt{\nu_M} + \pi\sqrt{\nu_M}
  \bigr)^2}{\pi\,(1-\pi)^2\,\nu_M + (1-\pi)\,\pi^2\,\nu_M}
  = 1 - \rho^2.
\end{equation}
This result holds for any
allocation ratio $\pi\in(0,1)$.
 
\textbf{Extension to local alternatives.}
Under the local alternative $\theta_0=cn^{-1/2}$ for a fixed constant
$c$, Lemma~S4 of \citet{ye2024covariate} establishes that
$E\{O_{ij}(\theta_0)\}\to 0$ and the asymptotic distributions of the
pseudo-outcomes converge to their $H_0$ limits. In particular,
$\nu_j(\theta_0)\to\nu_j(0)$ and $\rho_j(\theta_0)\to\rho_j(0)$ as
$n\to\infty$, so formula~\eqref{eq:VRR} holds asymptotically under
local alternatives as well.
 
We therefore conclude that under $H_0$ or local alternatives,
\begin{equation}\label{eq:VRR_main}
  \frac{\operatorname{Var}(\hat\theta_{CL})}{\operatorname{Var}(\hat\theta_L)}
  = \frac{\sigma_{CL}^{2}(\theta_0)}{\sigma_L^{2}(\theta_0)}
  \approx 1 - \rho^2,
\end{equation}

That is, the variance of the prognostic-score-adjusted log-HR estimator
is approximately $(1-\rho^2)$ times that of the unadjusted estimator.
This is the direct analogue of the PROCOVA result for continuous
endpoints \citep{schuler2022increasing}, with the key distinction that $\rho$
here measures the correlation between the prognostic score and the
survival pseudo-outcomes $O_{ij}(\theta_0)$, rather than the raw
outcome.  Higher prognostic value of the score, reflected in a larger
$|\rho|$, leads to greater variance reduction.  When the score carries
no prognostic information ($\rho=0$), then
$\sigma_{CL}^{2}=\sigma_L^{2}$ and no efficiency is lost.

\subsection{Event Number Reduction}\label{event-count-reduction}
 
The practical implications of the variance reduction
formula~\eqref{eq:VRR} extend naturally to trial planning.  Under the
log-rank framework, the number of events required to achieve power
$1-\beta$ at two-sided significance level $\alpha$ is proportional to
$\sigma_L^{2}(0)/\theta_0^{2}$
\citep{schoenfeld1981asymptotic}.  After prognostic score adjustment,
the required number of events scales with
$\sigma_{CL}^{2}(0)/\theta_0^{2}$.  Therefore:
\[
  \frac{d_{\text{adj}}}{d_{\text{unadj}}}
  = \frac{\sigma_{CL}^{2}(0)}{\sigma_L^{2}(0)}
  = 1-\rho^{2},
\] 
where $d_{\text{adj}}$ and $d_{\text{unadj}}$ denote the required
event counts with and without prognostic score adjustment,
respectively. Equivalently, the number of events that can be saved by
adopting the adjusted analysis is 
\begin{equation}\label{eq:events_saved}
  \text{Events saved} = \rho^2 \times d_{\mathrm{unadj}}.
\end{equation}
 
This result has direct implications for trial efficiency. If the
prognostic score explains, for example, 30\% of the variation in the
pseudo-outcomes ($\rho^2 \approx 0.30$), the required event count is
reduced by approximately 30\% relative to the unadjusted log-rank test.
Importantly, this reduction applies to the expected trial duration as well, since fewer events means a shorter
follow-up period is needed to accumulate the required information. 
If enrollment and study duration remain fixed, then this benefit materializes 
as increased power at the analysis stage. This increase in power also allows for more flexible trial designs, including more frequent interim monitoring without compromising
overall power.

The extension of all results in Sections~2 and~3 to stratified and
covariate-adaptive randomization is presented in Appendix \ref{variance-reduction-and-event-savings-under-stratified-randomization}, including
the stratified estimator $\hat\theta_{CSL}$, its asymptotic distribution,
variance estimator, and the corresponding variance reduction and event
savings formulas under prognostic score adjustment.

\subsection{Training the Prognostic Model from Historical Data}\label{training-the-prognostic-model-from-historical-data}
 
Equations~\eqref{eq:VRR_main} and~\eqref{eq:events_saved} show that
the strength of the prognostic score adjustment is governed by $\rho =
\operatorname{Corr}(\eta(X_i), O_{ij}(0))$, the arm-specific
correlation between the prognostic score and the log-rank
pseudo-outcomes. Maximising $|\rho|$ is therefore the natural training
criterion.

However, directly targeting $O_{ij}(0)$ is infeasible from a
historical control cohort: the pseudo-outcome
$\hat{O}_{ij} = \int_0^\tau \bar{Y}_{1-j}(t)/\bar{Y}(t)\,dM_{ij}(t)$
requires information from both treatment arms and thus cannot be formed
from controls alone.

\textbf{A computationally feasible training target.}
The martingale representation established in Section~3.1 provides a
natural resolution. Under $H_0$ and large $n$, we showed that
$O_{ij}(0) \to c_j(0)\cdot M_{ij}(\tau)$, where
$c_1(0)=1-\pi$, $c_0(0)=\pi$ are arm-specific constants and
$M_{ij}(\tau) = N_{ij}(\tau) - \int_0^\tau Y_{ij}(s)\,d\hat\Lambda(s)$
is the arm-$j$ martingale residual. Since correlations are invariant to
positive scaling,
\begin{equation}\label{eq:rho_M}
  \rho
  = \operatorname{Corr}\!\bigl(\eta(X_i),\,O_{ij}(0)\bigr)
  \approx \operatorname{Corr}\!\bigl(\eta(X_i),\,M_{ij}(\tau)\bigr).
\end{equation}
Moreover, $\operatorname{Corr}(\eta(X_i), M_{ij}(\tau))$ does not
depend on arm $j$ under randomization ($X_i \perp I_i$). It can
therefore be estimated from historical control data alone, by replacing
the trial arm-$j$ martingale with the analogous residual from the
external cohort. Specifically, define
\begin{equation}\label{eq:PS_M}
  \hat{M}_i^{\mathrm{ext}}(\tau)
  = \Delta_i - \hat{\Lambda}_0^{\mathrm{ext}}(\widetilde{T}_i),
\end{equation}
where $\hat{\Lambda}_0^{\mathrm{ext}}$ is the Nelson--Aalen estimator
of the baseline cumulative hazard fitted to the historical control
cohort, and $\widetilde{T}_i$ and $\Delta_i$ are the observed survival
time and event indicator for subject $i$ in the historical data. This
residual is positive on average for patients who experience the event
earlier than predicted under the estimated baseline hazard, and
negative for those who survive or are censored later. 
The prognostic model can therefore be trained
as \[
\eta(X_i) = E\!\bigl[\hat{M}_i^{\text{ext}}(\tau) \mid X_i\bigr],
\] and any supervised regression method can be used for estimation,
since $\hat{M}_i^{\text{ext}}(\tau)$ is a continuous response variable.
 
An important practical consequence of the model-free nature of the
\citet{ye2024covariate} framework is that the goodness of fit of the
prognostic model does not affect the validity of the primary inference.
Regardless of whether the prognostic model is correctly specified, or even
heavily overfit, the Type~I error of the adjusted test remains
controlled and the treatment effect estimator remains consistent.
Overfitting of the prognostic model could results in a
less-than-optimal $\rho$, but does not bias the estimator or inflate
the Type~I error rate. However, since an overfit $\hat\rho$ may
overestimate the true correlation, it leads to an overly optimistic
event savings prediction if used at the trial planning stage. We
therefore recommend applying a conservative correlation estimation
strategy at design time, such as the cross-validation-based procedure with discounting factors implemented as
described in the PROCOVA handbook \citep{PROCOVAhandbook}.

In summary, to implement the proposed method in practice, one (i) fits
a prognostic model $\eta(\cdot)$ on the historical control data,
targeting the cumulative martingale residual; (ii) evaluates $\hat{\eta}(X_i)$ for each concurrent trial
patient using their baseline covariates; and (iii) includes
$\hat{\eta}(X_i)$ as a baseline covariate in the adjusted log-rank
score $\hat{U}_{CL}(\vartheta)$ and HR estimator $\hat\theta_{CL}$
defined in Section~2.3. Steps (i)--(ii) use only the historical control
data and the trial's baseline measurements, introducing no additional
assumptions into the primary analysis.

\textbf{Alternative training target: predicted survival probability.}
A second practical option is to train the prognostic score by predicting
the survival probability function $\hat{S}(\cdot)$ on the historical
control data. The motivation is that $\hat{M}_i^{\text{ext}}(\tau)$
and $\hat{S}(\cdot)$ are both related to
$\hat{\Lambda}_0^{\text{ext}}(\widetilde{T}_i)$ and therefore carry
the similar prognostic information in terms of rank correlation
with the pseudo-outcomes. 
Predicting the survival probability therefore constitutes a reasonable
and often more interpretable surrogate. In particular,
survival probability can be predicted using off-the-shelf tools such
as random survival forests, which simultaneously estimate the full survival curve.
In our simulation studies we include
results from both training targets and confirm that the two perform
comparably across most scenarios.

\section{Simulations}\label{simulations}
 
We conduct a simulation study to evaluate the finite-sample operating
characteristics of the proposed method across a range of scenarios that
vary in the quality and relevance of the historical control data. The
study examines these properties: Type~I error control, power gain from
prognostic score adjustment when the historical data are informative,
and robustness when they are uninformative or misspecified. By our
theory, the adjusted estimator should be nearly unbiased and maintain
valid Type~I error in all scenarios, while achieving variance reduction
whenever the prognostic score correlates with the concurrent survival
outcomes.
 
\subsection{Setup}\label{setup}
 
In most scenarios, the conditional hazard for the target trial
follows a Cox model
\[
\lambda_j(t \mid X_1, X_2)
= h_0 \exp\!\bigl(0.8 + \theta j + \beta_1 X_1|X_2|
  - \beta_2(X_2 - 0.5)^2\bigr), \quad j = 0, 1.
\]
Censoring times follow an exponential distribution with rate 0.02,
independently of the event time and baseline covariates. Three baseline
covariates are observed in both the concurrent trial and the historical
control data: $X_1 \sim \operatorname{Bern}(0.5)$,
$X_2 \sim \mathcal{N}(0, 1)$, and $X_3 \sim \mathcal{N}(0,1)$.
Only $X_1$ and $X_2$ enter the data-generating mechanism; $X_3$ is a
noise variable. The baseline hazard is $h_0 = 0.08$,
$\beta_1 = \log(1.8)$, $\beta_2 = \log(3)$. The parameter $\theta$
represents the conditional treatment effect on the log-hazard scale
and is numerically different from the target unconditional log-HR $\theta_0$
due to non-collapsibility, though both equal zero under the null. In the
null scenario $\theta = 0$; in the efficacy scenario
$\theta = \log(0.6)$.
 
The simulation data are generated under seven scenarios designed
to probe different dimensions of the method's robustness:
 
\noindent \textbf{Case~I: Ideal external data.} The historical
control hazard is identical to the concurrent control hazard:
$\tilde\lambda_0(t \mid X_1, X_2) = \lambda_0(t \mid X_1, X_2)$.
This represents the most favorable case.
 
\noindent \textbf{Case~II: Different baseline hazard and model.}
The historical control follows
$\tilde\lambda_0(t \mid X_1, X_2) = 0.05\exp(0.8 + 0.2X_2)$,
with a different baseline rate and a different covariate structure.
The prognostic score captures some but limited information about
concurrent outcomes.
 
\noindent \textbf{Case~III: Missing key covariate.}
The historical control hazard is the same as in Case~I, but $X_2$ is
unavailable in the historical dataset and is excluded from prognostic
model training. The score is trained on $X_1$ and $X_3$ alone, severely
limiting its predictive value.
 
\noindent \textbf{Case~IV: Completely uninformative historical data.}
The historical control hazard is constant:
$\tilde\lambda_0(t) = h_0\exp(0.8)$, with no dependence on covariates.
The prognostic score carries no useful information about concurrent
outcomes.
 
\noindent \textbf{Case~V: Different distributional family.}
Historical control survival times follow a log-normal AFT model:
$\log T = 1 + \beta_1 X_1 X_2 + \beta_2(X_2 - X_1)^2 + 0.7\varepsilon$,
where $\varepsilon \sim \mathcal{N}(0,1)$. The distribution differs
from the target trial but the covariate structure is similar, so
moderate prognostic signal can be captured.
 
\noindent \textbf{Case~VI: Log-normal AFT for target trial.}
Both the target and historical outcomes follow a log-normal AFT
model:
$\log T = 1 - \theta j - \beta_1 X_1|X_2| + \beta_2(X_2-X_1)^2
+ \delta\varepsilon$, with $\delta = 0.5$ for the concurrent study
and $\delta = 0.7$ for the historical control. Notably, the proportional hazards
do not hold in this setting.

\noindent \textbf{Case~VII: Piecewise hazards.}
The target trial follows a two-piece exponential model with a change point at $\tau = 5$, where covariate effects vary across the two intervals $I_1 = [0,\tau)$ and $I_2 = [\tau,\infty)$. Specifically, for $k = 1,2$,
$$
\lambda_j(t \mid X_1, X_2) = h_0 \exp\!\Bigl(0.8 + c_{A,k}\,\theta j + c_{1,k}\,\beta_1 X_1|X_2| - c_{2,k}\,\beta_2 (X_2 - 0.5)^2 \Bigr), \quad t \in I_k,
$$
where $(c_{A,1}, c_{A,2}) = (0.8, 1.2)$, $(c_{1,1}, c_{1,2}) = (0.8, 1.2)$, and $(c_{2,1}, c_{2,2}) = (1.2, 0.8)$. This induces non-proportional hazards through time-varying effects in the target trial, while the historical control data are generated as in Case~I.

The significance level is $\alpha = 5\%$, the target allocation
proportion $\pi = 0.5$, and the target trial sample size is
$n = 200$ or $400$. The historical control sample size is fixed at 300.
Noteworthy, it is generated once and held fixed across
all 10,000 simulation replicates, mimicking the real-world scenario
in which historical/external data are collected prior to the trial and remain
fixed at the time of analysis.
 
\subsection{Analysis and Results}\label{analysis-and-results}
 
The unadjusted marginal HR and its variance are obtained
from a Cox model with treatment as the sole covariate. For the proposed
method, the prognostic score is trained on the historical control data
following equation~\eqref{eq:PS_M} using regression random forests
\citep{breiman2001random}, targeting the cumulative martingale residual.
The fitted score $\hat\eta(X_i)$ is then included as the sole adjustment
covariate in $\hat\theta_{CL}$. 
Hypothesis testing uses the covariate-adjusted log-rank statistic
$T_{CL}$, which is asymptotically equivalent to the Wald statistic
based on $\hat\theta_{CL}$ and $\widehat{\operatorname{Var}}(\hat\theta_{CL})$.
The external data is used in whole for model fitting.
 
\begin{figure}[ht]
\centering
\includegraphics[width=0.96\textwidth, height=0.82\textheight, keepaspectratio]{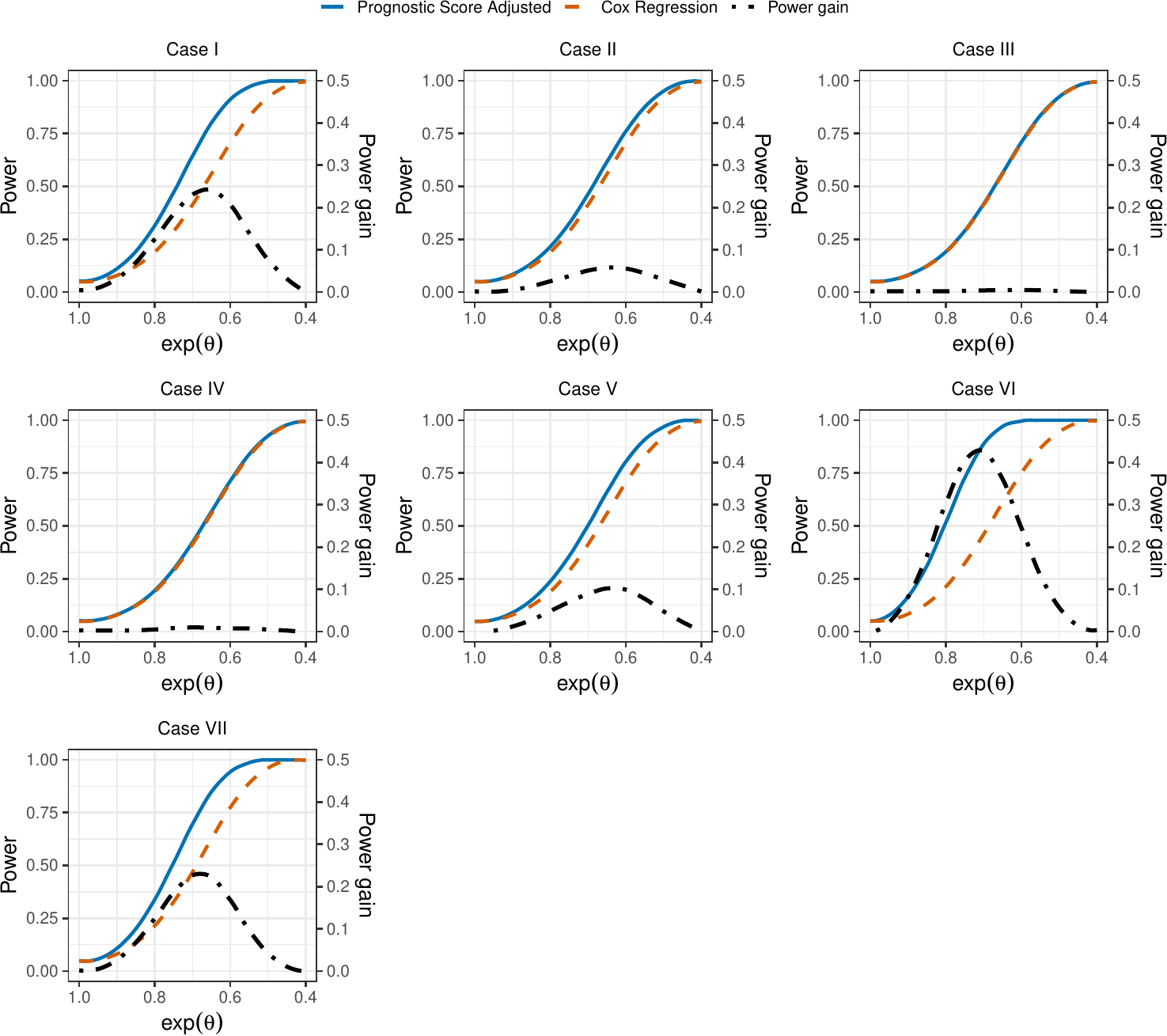}
\caption{Power curves and power gain across seven simulation scenarios
at $n = 400$. Each panel shows the empirical power from the unadjusted
Cox regression analysis (dashed) and the prognostic-score-adjusted
analysis (solid), plotted on the left axis. The absolute power gain
(dash-dot) is plotted on the right axis.}
\label{fig:PowerFig}
\end{figure}

The simulation results are reported in Table~\ref{tb:SimuI} and
Figure~\ref{fig:PowerFig}. We highlight the following key findings.

\textbf{Bias and Type~I error control.}
The absolute bias, defined as the difference in unconditional log-HR between
the adjusted and unadjusted estimates, is negligible across all cases,
scenarios, and sample sizes (column ``Bias'' $\approx 0$ throughout). 
This confirms that the prognostic score
adjustment introduces no bias in the treatment effect estimator,
regardless of whether the historical data are informative, misspecified,
or entirely irrelevant. Moreover, the rejection probability under the null
(``Pr.~Reject H0: Cov-Adj'' with $\theta = 0$) is consistently close
to the nominal 5\% level across all seven cases, show no inflation relative to the unadjusted test.

\textbf{Power gain and variance reduction.}
When the prognostic score is correlated with the concurrent survival
outcomes (Cases~I, V, VI, and~VII), meaningful power gains are observed.
In Case~I, for example, power increases from approximately 43\% to 65\% at $n=200$
and from 71\% to 91\% at $n=400$. 
The power curves in Figure~\ref{fig:PowerFig} display the full picture across the range
of $\exp(\theta)$: Cases~I, VI, and~VII show the largest and
most sustained gains, while Cases~III and~IV produce power curves
that are essentially indistinguishable from the unadjusted analysis.
Table~\ref{tb:SimuI} further reports the empirical variance reduction
ratio $\text{Var}(\hat\theta_{CL})/\text{Var}(\hat\theta_{L})$, alongside $1 - \hat\rho^2$, the approximation from
equation~\eqref{eq:VRR_main}, where $\hat\rho$ is the correlation between
the fitted prognostic score and the target trial martingale proxy
$\hat{M}(\tau)$. The two quantities are in close agreement across all
cases, which validates the analytical approximation in~\eqref{eq:VRR_main}
and confirms that the martingale residual is a good proxy for the
pseudo-outcomes when computing $\rho$. 
 
\textbf{Variance estimator validity.} The columns MSE (mean of theoretical
standard error from equation~\ref{eq:VarEq}) and MCSD (Monte Carlo
standard deviation of $\hat\theta_{CL}$ across 10,000 replicates) are
in close agreement across all settings, confirming that the
variance estimator is accurate in finite samples.

Additional simulation results, including analyses with the survival
probability as the training target (using R package \texttt{randomForestSRC}) and analyses incorporating observed
baseline covariates in addition to the prognostic score, are reported
in Appendix Table~\ref{tb:SimuII} (Section~\ref{additional-simulation-results}).

The main conclusions are: (i)~training on the survival probability
$\hat{S}(t)$ performs comparably to training on $\hat{M}^{\text{ext}}$;
and (ii)~including observed baseline covariates alongside the prognostic
score provides additional power gain in Cases~II--V, where the external
information is only partially captured by the score. This supports the
recommendation that in practice, the prognostic score should be included
\emph{in addition to} observed baseline covariates, not as a replacement.
 
\begingroup\fontsize{9}{11}\selectfont
 
\begin{longtable}[t]{cccccccccc}
\caption{\label{tab:HRinfoBorrow}\label{tb:SimuI}Simulation results for the proposed covariate-adjusted HR estimator across seven external control scenarios and two target RCT trial sample sizes ($n = 200, 400$). Bias: absolute difference in log-HR between the adjusted and unadjusted estimates. Pr.~Reject~H0: Cox (respectively, Cov-Adj): empirical rejection rate for the unadjusted (respectively, adjusted) test; equals Type~I error under the null scenario and power under the efficacy scenario. MSE: mean of the estimated standard error from equation~(\ref{eq:VarEq}). MCSD: Monte Carlo standard deviation of $\hat\theta_{CL}$ across 10,000 replicates. $\text{Var}(\hat\theta_{CL})/\text{Var}(\hat\theta_{L})$: empirical variance reduction ratio. $\hat\rho$: sample correlation between the fitted prognostic score and the concurrent-trial martingale proxy $\hat{M}(\tau)$. $1-\hat\rho^2$: approximation from equation~(\ref{eq:VRR_main}).}\\
\toprule
 & $n$ & Bias & Pr.\ Rej.\ H0: Cox & Pr.\ Rej.\ H0: Cov-Adj & MSE & MCSD & $\text{Var}(\hat\theta_{CL})/\text{Var}(\hat\theta_{L})$ & $\hat\rho$ & $1-\hat\rho^2$\\
\midrule
\addlinespace[0.3em]
\multicolumn{10}{l}{\textbf{Case I: Ideal external data}}\\
\hspace{1em} & 200 & 0.001 & 0.051 & 0.046 & 0.123 & 0.123 & 0.539 & 0.679 & 0.538\\
\cmidrule{2-10}\nopagebreak
\hspace{1em}\multirow{-2}{*}{\centering\arraybackslash Null} & 400 & 0.000 & 0.047 & 0.051 & 0.087 & 0.087 & 0.536 & 0.681 & 0.535\\
\cmidrule{1-10}\pagebreak[0]
\hspace{1em} & 200 & 0.001 & 0.426 & 0.645 & 0.131 & 0.131 & 0.571 & 0.648 & 0.578\\
\cmidrule{2-10}\nopagebreak
\hspace{1em}\multirow{-2}{*}{\centering\arraybackslash Efficacy} & 400 & 0.000 & 0.707 & 0.914 & 0.092 & 0.093 & 0.568 & 0.651 & 0.575\\
\cmidrule{1-10}\pagebreak[0]
\addlinespace[0.3em]
\multicolumn{10}{l}{\textbf{Case II: Different baseline hazard and model}}\\
\hspace{1em} & 200 & 0.000 & 0.051 & 0.052 & 0.156 & 0.157 & 0.862 & 0.367 & 0.860\\
\cmidrule{2-10}\nopagebreak
\hspace{1em}\multirow{-2}{*}{\centering\arraybackslash Null} & 400 & 0.000 & 0.047 & 0.050 & 0.110 & 0.109 & 0.861 & 0.370 & 0.860\\
\cmidrule{1-10}\pagebreak[0]
\hspace{1em} & 200 & 0.000 & 0.426 & 0.474 & 0.161 & 0.163 & 0.874 & 0.346 & 0.875\\
\cmidrule{2-10}\nopagebreak
\hspace{1em}\multirow{-2}{*}{\centering\arraybackslash Efficacy} & 400 & 0.000 & 0.707 & 0.759 & 0.114 & 0.114 & 0.873 & 0.349 & 0.875\\
\cmidrule{1-10}\pagebreak[0]
\addlinespace[0.3em]
\multicolumn{10}{l}{\textbf{Case III: Missing key covariate in historical data}}\\
\hspace{1em} & 200 & 0.000 & 0.051 & 0.052 & 0.167 & 0.169 & 0.993 & 0.050 & 0.993\\
\cmidrule{2-10}\nopagebreak
\hspace{1em}\multirow{-2}{*}{\centering\arraybackslash Null} & 400 & 0.000 & 0.047 & 0.049 & 0.118 & 0.117 & 0.995 & 0.049 & 0.995\\
\cmidrule{1-10}\pagebreak[0]
\hspace{1em} & 200 & 0.001 & 0.426 & 0.431 & 0.172 & 0.174 & 0.993 & 0.049 & 0.993\\
\cmidrule{2-10}\nopagebreak
\hspace{1em}\multirow{-2}{*}{\centering\arraybackslash Efficacy} & 400 & 0.001 & 0.707 & 0.712 & 0.121 & 0.121 & 0.995 & 0.048 & 0.995\\
\cmidrule{1-10}\pagebreak[0]
\addlinespace[0.3em]
\multicolumn{10}{l}{\textbf{Case IV: Completely uninformative historical data}}\\
\hspace{1em} & 200 & 0.000 & 0.051 & 0.053 & 0.166 & 0.168 & 0.979 & 0.122 & 0.977\\
\cmidrule{2-10}\nopagebreak
\hspace{1em}\multirow{-2}{*}{\centering\arraybackslash Null} & 400 & 0.000 & 0.047 & 0.050 & 0.117 & 0.116 & 0.981 & 0.126 & 0.980\\
\cmidrule{1-10}\pagebreak[0]
\hspace{1em} & 200 & 0.001 & 0.426 & 0.437 & 0.171 & 0.174 & 0.981 & 0.110 & 0.980\\
\cmidrule{2-10}\nopagebreak
\hspace{1em}\multirow{-2}{*}{\centering\arraybackslash Efficacy} & 400 & 0.000 & 0.707 & 0.714 & 0.121 & 0.121 & 0.984 & 0.114 & 0.983\\
\cmidrule{1-10}\pagebreak[0]
\addlinespace[0.3em]
\multicolumn{10}{l}{\textbf{Case V: Different distributional family (log-normal AFT)}}\\
\hspace{1em} & 200 & 0.001 & 0.051 & 0.051 & 0.147 & 0.149 & 0.773 & 0.474 & 0.772\\
\cmidrule{2-10}\nopagebreak
\hspace{1em}\multirow{-2}{*}{\centering\arraybackslash Null} & 400 & 0.001 & 0.047 & 0.047 & 0.104 & 0.104 & 0.772 & 0.477 & 0.771\\
\cmidrule{1-10}\pagebreak[0]
\hspace{1em} & 200 & 0.001 & 0.426 & 0.515 & 0.153 & 0.155 & 0.788 & 0.454 & 0.790\\
\cmidrule{2-10}\nopagebreak
\hspace{1em}\multirow{-2}{*}{\centering\arraybackslash Efficacy} & 400 & 0.000 & 0.707 & 0.807 & 0.108 & 0.108 & 0.787 & 0.456 & 0.790\\
\cmidrule{1-10}\pagebreak[0]
\addlinespace[0.3em]
\multicolumn{10}{l}{\textbf{Case VI: Log-normal AFT in target trial}}\\
\hspace{1em} & 200 & 0.003 & 0.053 & 0.048 & 0.094 & 0.096 & 0.320 & 0.824 & 0.321\\
\cmidrule{2-10}\nopagebreak
\hspace{1em}\multirow{-2}{*}{\centering\arraybackslash Null} & 400 & 0.000 & 0.048 & 0.051 & 0.067 & 0.067 & 0.319 & 0.825 & 0.320\\
\cmidrule{1-10}\pagebreak[0]
\hspace{1em} & 200 & 0.002 & 0.477 & 0.890 & 0.101 & 0.103 & 0.348 & 0.801 & 0.358\\
\cmidrule{2-10}\nopagebreak
\hspace{1em}\multirow{-2}{*}{\centering\arraybackslash Efficacy} & 400 & 0.001 & 0.750 & 0.994 & 0.071 & 0.072 & 0.347 & 0.802 & 0.357\\
\cmidrule{1-10}\pagebreak[0]
\addlinespace[0.3em]
\multicolumn{10}{l}{\textbf{Case VII: Piecewise hazards}}\\
\hspace{1em} & 200 & 0.001 & 0.053 & 0.052 & 0.123 & 0.124 & 0.555 & 0.667 & 0.553\\
\cmidrule{2-10}\nopagebreak
\hspace{1em}\multirow{-2}{*}{\centering\arraybackslash Null} & 400 & 0.000 & 0.046 & 0.049 & 0.086 & 0.086 & 0.553 & 0.669 & 0.552\\
\cmidrule{1-10}\pagebreak[0]
\hspace{1em} & 200 & 0.001 & 0.488 & 0.711 & 0.131 & 0.133 & 0.592 & 0.630 & 0.600\\
\cmidrule{2-10}\nopagebreak
\hspace{1em}\multirow{-2}{*}{\centering\arraybackslash Efficacy} & 400 & 0.000 & 0.775 & 0.944 & 0.092 & 0.092 & 0.589 & 0.633 & 0.599\\
\bottomrule
\end{longtable}
\endgroup{}

\section{A Real Data Application}\label{a-real-data-application}

\subsection{Study Descriptions and Data}

We illustrate the proposed method using two clinical trial datasets in
second-line metastatic colorectal cancer (mCRC) studies that share similar control population.

\textbf{External control data.}
The prognostic model is trained on the control arm of a Phase~III
randomized trial (NCT00561470) evaluating aflibercept plus FOLFIRI versus
placebo plus FOLFIRI in patients with mCRC previously treated with an
oxaliplatin-based regimen \citep{van2012addition}. We use the placebo
plus FOLFIRI arm ($n = 604$ after excluding patients with missing
outcomes from the randomized $n = 614$). The median OS in this arm
is 12.1~months (95\% CI: 11.0--13.1), consistent with the published
report.

\textbf{Target Phase III data.}
The primary analysis is applied to study NCT00339183, which is an open-label,
Phase~III randomized trial evaluating panitumumab plus FOLFIRI versus
FOLFIRI alone as second-line therapy for mCRC \citep{peeters2010randomized, peeters2014final}.
In this work, we target on OS as the primary endpoint. The trial enrolled 1,186
patients: 591 in the panitumumab plus FOLFIRI arm and 595 in the
FOLFIRI-alone arm, with 419 and 435 OS events, respectively. KRAS
mutation status was assessed at baseline and available for 91\% of patients: 597 KRAS wild-type
patients (303 panitumumab, 294 control) and 486 KRAS mutant patients
(238 panitumumab, 248 control). Median OS in the FOLFIRI-alone arm
is 12.2~months overall, 12.5~months in the KRAS wild-type subgroup,
and 11.1~months in the KRAS mutant subgroup.

The two control populations are clinically comparable: both received
FOLFIRI as second-line therapy for mCRC following prior
treatment, and the median OS in the respective
control arms is nearly identical (12.1 versus 12.2~months). This
similarity supports the use of the external control data to train the
prognostic model for the target trial.

\subsection{Prognostic Model Training}

Fifteen baseline variables are shared between the two datasets and
used as predictors: ECOG performance status, prior
bevacizumab use (yes/no), primary tumor site, baseline LDH
($\geq 1.5\times$ ULN, binary), and eleven
laboratory markers (alkaline phosphatase, albumin, neutrophils,
bilirubin, hemoglobin, alanine aminotransferase, platelet count,
aspartate aminotransferase, potassium, sodium, and calcium). KRAS mutation
status, an established predictive
biomarker for anti-EGFR therapy in mCRC, is not recorded in the
external control dataset and is therefore excluded for model training.
Missing covariate values are sparse in both datasets (at most 4.3\%
in the external data and 3.5\% in the target trial data) and are
imputed using \texttt{missForest} \citep{stekhoven2012missforest} before model
fitting.

The cumulative martingale residual $\hat{M}_i^{\text{ext}}(\tau)$
is computed on the external control arm following equation~\eqref{eq:PS_M}.
Following the simulation studies, a regression random forest is fitted
to predict $\hat{M}_i^{\text{ext}}(\tau)$ from the 15 baseline
covariates using the \texttt{ranger} package \citep{wright2017ranger}
with 2,000 trees and maximum depth~5. We also fitted with BART \citep{chipman2010bart} and
SuperLearner \citep{van2007super} as alternative learning methods. 
The resulting variance reduction estimates were consistent with
those from the random forest, so only the random forest results are
reported. The fitted model is applied to each patient in the target
trial to produce the predicted prognostic score $\hat\eta(X_i)$.

For simplicity, the external control data are used in their entirety
for model fitting without cross-validation or a held-out validation
set. This is because this application is purely
illustrative. In practice, particularly when the
prognostic score is used to plan a future trial, a cross-validation
strategy is recommended to obtain a conservative estimate of $\rho$
and avoid overstating the expected efficiency gain, as discussed in
Section~3.3.

\subsection{Results}

The covariate-adjusted HR estimator are estimated
as described in the previous sections, with the pre-trained prognostic
score as the sole adjustment covariate. We report results for the
overall population and for the two KRAS subgroups. All analyses are
unadjusted for multiplicity and are intended as illustrations of the
method rather than confirmatory tests.

Table~\ref{tab:RealData} summarizes the results. The covariate-adjusted HR estimates 
are numerically close to their unadjusted counterparts in all three analyses ($0.899$ versus
$0.907$ overall; $0.866$ versus $0.865$ in the KRAS wild-type
subgroup; $0.922$ versus $0.929$ in the KRAS mutant
subgroup). This is consistent with the theoretical guarantee that the
adjustment does not introduce bias to the unconditional HR estimand.

In the overall population, the standard error of the log-HR decreases
from 0.069 to 0.062 after prognostic score adjustment, corresponding
to a variance reduction of approximately 17.6\%. In the KRAS wild-type subgroup, the standard error decreases
from 0.099 to 0.092, yielding a 14.2\% variance reduction. In the KRAS mutant subgroup, the reduction
gets even larger at approximately 23.3\%, which suggests that
the prognostic score captures substantial individual-level survival
variability in this group, despite the absence of KRAS from the
training data.

Overall, this application demonstrates that prognostic score 
adjusted analysis yields improved precision and 
equivalently, increased power of the study. 
When applied at the design stage, the resulting variance reduction 
can be translated into savings in the required number of events and lead to shorter study duration. 
For example, if a 20\% variance reduction is anticipated
from prognostic score adjustment, then approximately the same power would have been achievable with about
20\% fewer events required, potentially saving months  in study duration depending on the enrollment speed and event rate. This would accelerate development timelines nad bringing effective medicines to patients sooner by efficiently leverage external information.
Therefore, the pre-trained prognostic score using external data offer a robust and efficient approach to borrowing external/historical information to improve trial efficiency in confirmatory settings, without inflating Type-I-error or introducing bias.

\begin{table}[h!]
\centering
\caption{Covariate-adjusted versus unadjusted analysis of OS for
target trial (NCT00339183), using the external control data from NCT00561470 trial for prognostic model training.
Log-HR: log hazard ratio (panitumumab plus FOLFIRI versus FOLFIRI alone).
SE: standard error of the log-HR. HR: hazard ratio.
Var.\ reduction: $1 - (\text{SE}_{\text{adj}}/\text{SE}_{\text{unadj}})^2$.
p-values are one-sided.}
\label{tab:RealData}
\renewcommand{\arraystretch}{1.35}
\begin{tabular}{llcccccc}
\toprule
Population & Method & $n$ & Events & Log-HR & SE & HR & $p$-value \\
\midrule
\multirow{2}{*}{Overall}
  & Unadjusted & 1186 & 854 & $-0.098$ & $0.069$ & $0.907$ & $0.076$ \\
  & Adjusted   & 1186 & 854 & $-0.107$ & $0.062$ & $0.899$ & $0.044$ \\
\multicolumn{8}{r}{\textit{Variance reduction: 17.6\%}} \\
\midrule
\multirow{2}{*}{KRAS wild-type}
  & Unadjusted & 597 & 407 & $-0.145$ & $0.099$ & $0.865$ & $0.072$ \\
  & Adjusted   & 597 & 407 & $-0.144$ & $0.092$ & $0.866$ & $0.058$ \\
\multicolumn{8}{r}{\textit{Variance reduction: 14.2\%}} \\
\midrule
\multirow{2}{*}{KRAS mutant}
  & Unadjusted & 486 & 374 & $-0.074$ & $0.104$ & $0.929$ & $0.236$ \\
  & Adjusted   & 486 & 374 & $-0.081$ & $0.091$ & $0.922$ & $0.187$ \\
\multicolumn{8}{r}{\textit{Variance reduction: 23.3\%}} \\
\bottomrule
\multicolumn{8}{l}{}
\end{tabular}
\end{table}

\section{Discussion}\label{discussions}

We have proposed a risk-free framework for borrowing information
from external/historical data to improve the efficiency of unconditional HR estimation in randomized clinical trials with
time-to-event endpoints. The proposed method combines two ideas in a novel way. The nonparametric covariate-adjusted log-rank test and HR estimator of
\citet{ye2024covariate} provide a robust analysis framework, while the PROCOVA \citep{schuler2022increasing} offers a principled mechanism for encoding historical information into a single baseline covariate. This strategy avoids direct inclusion of subject-level external data into
the primary analysis, thereby mitigating the risk of Type I error inflation.

The central methodological challenge addressed in this paper is the
extension of prognostic-score adjustment to time-to-event endpoints
while preserving the unconditional hazard ratio as the primary
estimand. Conventional Cox covariate adjustment fails to meet this
challenge because the hazard ratio is non-collapsible, adding
covariates to a Cox model shifts the estimand from the unconditional to
the conditional HR, even under correct model specification and
randomization. By embedding an externally trained prognostic score
within the model-free covariate-adjusted framework of
\citet{ye2024covariate}, the proposed method achieves efficiency gain
without altering the estimand and without relying on any assumption
about the hazard or censoring distribution. This resolves the key
obstacle that has previously limited the extension of PROCOVA-style
methods to survival endpoints, and provides a principled and
audit-friendly pathway for incorporating high-dimensional, 
machine-learning-derived prognostic information into the primary 
survival analysis of a confirmatory trial--- something that has 
previously been difficult to justify under strict frequentist error control.

A structural feature of the proposed framework deserves emphasis. 
The two stages of the method are entirely decoupled: the prognostic 
model is trained on external data using any method the 
analyst chooses, and it enters the primary analysis solely as a 
fixed baseline covariate. The inferential stage is then a standard, 
pre-specified, model-free analysis whose validity is guaranteed by 
randomization, not by any property of the prognostic model. This 
separation means that advances in machine learning and the growing 
availability of real-world data and historical trial databases 
translate directly into efficiency gains for confirmatory RCTs, 
without requiring any modification to the regulatory analysis or 
any relaxation of Type~I error guarantees. A richer external 
dataset or a better prognostic model yields a larger $\rho$ and 
therefore a larger efficiency gain, but for the worst case, one just yields  
$\rho \approx 0$ with no harm done. The framework therefore provides
a principled, scalable approach to incorporating
high-dimensional prognostic information into the primary survival
analysis of a confirmatory trial.

From a design perspective, the variance reduction formula
$\sigma_{CL}^2/\sigma_L^2 \approx 1 - \rho^2$ provides a simple and actionable planning tool.
Once $\rho$ is estimated from the historical data via the martingale
residual proxy described in Section~3.3, the
analyst can quantify the expected efficiency gain before the trial
begins. A trial that would require $d_{\text{unadj}}$ events under the
unadjusted log-rank test needs only approximately
$(1-\rho^2) \times d_{\text{unadj}}$ events under the adjusted analysis
at the same power. This reduction can be realized as a smaller required
sample size, a shorter follow-up period, greater power at fixed design
parameters, a greater number of planned interim analyses, or a combination of these. 
The essence is that this reduced event requirement allows more flexible trial designs in practice.

Based on the theoretical results and simulation evidence in the Appendix Table~\ref{tb:SimuII}, we recommend
including the prognostic score in addition to other observed baseline
covariates, not as a replacement. When the external data are informative,
especially with complicated non-linear structure,
the prognostic score captures external prognostic information that the
baseline covariates alone cannot. When the external data are not
informative, the baseline covariates still have the opportunity to provide efficiency gain from
within-trial prognostic information. This combined strategy ensures that
both internal and external information is used,
with no risk to the integrity of the primary inference.

When the trial uses stratified randomization, this work
can naturally be extended with a parallel structure as simple randomization.
We can apply the exact same prognostic model training and adjustment approach 
to leverage the external information to increase the precision of the estimator. 
See details in the Appendix~\ref{variance-reduction-and-event-savings-under-stratified-randomization}. 
When the stratification variable
is associated with the prognostic score, some of the score's variation
is already absorbed by the strata, and the achievable efficiency gain
from the score is somewhat smaller than under simple randomization.
The two sources of efficiency gain are complementary: stratification
accounts for between-stratum outcome variation, while the prognostic
score addresses within-stratum variation that stratification does not
capture.
 
Throughout the simulation and real data application, 
we have described the external dataset as a
historical control cohort, which is the most common practical scenario
and the most conservative choice. A control dataset alone ensures
that the learned prognostic score captures only prognostic effects,
unconfounded by any treatment-specific signal. However, if both arms of a
historical trial are available, one may train the prognostic model on
the combined dataset with the treatment indicator excluded, providing
there are no strong predictive variables.

A final contrast with Bayesian dynamic borrowing is worth emphasizing. Bayesian borrowing methods incorporate external outcomes directly into the likelihood or prior for the treatment effect, so their operating characteristics depend critically on whether the historical and current populations are exchangeable, which is a condition that cannot be verified prospectively. The proposed approach operates on an entirely different principle: external data enter the analysis only through a fixed baseline covariate computed before the trial begins, and the validity of the primary test and estimator is protected entirely by randomization. Any predictive content in the prognostic score translates into efficiency gain; if the score carries no signal, no harm is done. The framework therefore offers what Bayesian borrowing cannot: an unconditional guarantee of Type~I error control alongside the opportunity for substantial efficiency gains.

\section{Appendix}\label{appendix}

\subsection{Extension to Stratified Randomization}\label{variance-reduction-and-event-savings-under-stratified-randomization}

This appendix develops the parallel of Sections~2 and~3 for trials
that employ stratified or covariate-adaptive randomization. We follow
the framework of \citet{ye2024covariate} (Sections~4 and~S3 of their
paper) throughout.

\subsubsection{Setup and Conditions}\label{app:strat-setup}

Let $Z_i \in \mathcal{Z}$ denote a discrete stratification variable
with finitely many levels, observed before randomization. We replace
Condition~1 with the following.

\textbf{Condition A1} (Covariate-adaptive randomization;
\citealt{ye2024covariate}). The stratification variable $Z_i$ takes
values in a finite set $\mathcal{Z}$; conditional on
$(Z_1,\ldots,Z_n)$, the treatment assignments $(I_1,\ldots,I_n)$ are
jointly independent of all potential outcomes; $\mathbb{E}(I_i \mid
Z_1,\ldots,Z_n) = \pi$ for all $i$; and $n_z^{-1}D_n(z) = o_p(1)$
for every $z \in \mathcal{Z}$, where $n_z = \#\{i : Z_i = z\}$ and
$D_n(z) = \sum_{i:\,Z_i=z}(I_i - \pi)$ is the within-stratum
treatment imbalance.

Condition~A1 is satisfied by stratified permuted block randomization
and Pocock--Simon minimization. Simple
randomization also satisfies Condition~A1. Assumption~1 (non-informative
censoring) is maintained as stated in Section~2.1, with $I_i$
replaced by $(I_i, Z_i)$.

Let $p_z = \Pr(Z_i = z)$, $n_{zj} = \#\{i : Z_i = z,\, I_i = j\}$,
and $n_z = n_{z0} + n_{z1}$. By Condition~A1,
$n_{zj}/n_z \to \pi$ in probability for every $z$.

\subsubsection{Stratified Estimator and Asymptotic Distribution}\label{app:strat-asym}

For stratum $z$, define stratum-specific at-risk processes
$\bar{Y}_{zj}(t) = n^{-1}\sum_{i:\,Z_i=z} Y_{ij}(t)$ and the
stratum-specific pseudo-outcomes
\begin{equation}\label{eq:PseudoO_strat}
\hat{O}_{zij}(\vartheta) = \int_0^\tau
\frac{e^{(1-j)\vartheta}\bar{Y}_{z,1-j}(t)}
{e^\vartheta\bar{Y}_{z1}(t) + \bar{Y}_{z0}(t)}
\left[dN_{zij}(t) - Y_{zij}(t)
\frac{e^{j\vartheta}\,d\bar{N}_z(t)}
{e^\vartheta\bar{Y}_{z1}(t) + \bar{Y}_{z0}(t)}\right],
\end{equation}
where $N_{zij}(t) = \mathbf{1}(\widetilde{T}_i \le t,\,\Delta_i=1,\,
I_i=j,\,Z_i=z)$ and $\bar{N}_z(t) = n^{-1}\sum_{i:\,Z_i=z}N_i(t)$.
At $\vartheta = 0$ these reduce to the stratum-specific analogues of
$\hat{O}_{ij}$ in equation~\eqref{eq:PseudoO}. The \textbf{stratified
unadjusted score} is
\[
\hat{U}_{SL}(\vartheta) = \frac{1}{n}\sum_z\sum_{i:\,Z_i=z}
\int_0^\tau\!\left(I_i - \frac{e^\vartheta\bar{Y}_{z1}(t)}
{e^\vartheta\bar{Y}_{z1}(t)+\bar{Y}_{z0}(t)}\right)dN_i(t),
\]
and $\hat\theta_{SL}$ solves $\hat{U}_{SL}(\vartheta) = 0$.

Within-stratum regression coefficients (pooled across strata) are
\[
\hat\gamma_j(\vartheta) =
\widehat{\mathbb{E}\{\operatorname{var}(X_i\mid Z_i)\}}^{-1}
\cdot \frac{1}{n}\sum_z\sum_{i:\,Z_i=z,\,I_i=j}
(X_i - \bar{X}_z)\hat{O}_{zij}(\vartheta), \quad j = 0,1,
\]
where $\bar{X}_z = n_z^{-1}\sum_{i:\,Z_i=z} X_i$ is the stratum-$z$
sample mean and
$\widehat{\mathbb{E}\{\operatorname{var}(X_i\mid Z_i)\}} =
n^{-1}\sum_z\sum_{i:\,Z_i=z}(X_i-\bar{X}_z)(X_i-\bar{X}_z)^\top$
is the pooled within-stratum sample covariance. The
\textbf{covariate-adjusted stratified score} is
\begin{equation}\label{eq:CSL_Score}
\hat{U}_{CSL}(\vartheta) = \hat{U}_{SL}(\vartheta)
- \frac{1}{n}\sum_z\sum_{i:\,Z_i=z}
\bigl\{I_i(X_i-\bar{X}_z)^\top\hat\gamma_1(\hat\theta_{SL})
- (1-I_i)(X_i-\bar{X}_z)^\top\hat\gamma_0(\hat\theta_{SL})\bigr\},
\end{equation}
and $\hat\theta_{CSL}$ is defined by solving
$\hat{U}_{CSL}(\vartheta) = 0$. As in the simple randomization case,
the augmentation is evaluated at $\hat\theta_{SL}$, ensuring
$\partial\hat{U}_{CSL}/\partial\vartheta =
\partial\hat{U}_{SL}/\partial\vartheta$.

By the M-estimation argument of Section~2.3.2, applied stratum by
stratum, and consistent with Theorem~S3 of \citet{ye2024covariate}:
\begin{equation}\label{eq:AN_strat}
n^{1/2}(\hat\theta_{CSL} - \theta_0) \xrightarrow{d}
N\!\left(0,\;\frac{\sigma^2_{CSL}(\theta_0)}
{\{\sigma^2_{SL}(\theta_0)\}^2}\right),
\end{equation}
where
\begin{align}
\sigma^2_{SL}(\theta_0) &= \sum_{z} p_z
\bigl\{\pi\nu_{z1}(\theta_0) + (1-\pi)\nu_{z0}(\theta_0)\bigr\},
\label{eq:sigSL}\\
\sigma^2_{CSL}(\theta_0) &= \sigma^2_{SL}(\theta_0)
- \pi(1-\pi)(\gamma_1(\theta_0)+\gamma_0(\theta_0))^\top
\mathbb{E}\!\bigl\{\operatorname{var}(X_i\mid Z_i)\bigr\}
(\gamma_1(\theta_0)+\gamma_0(\theta_0)),
\label{eq:sigCSL}
\end{align}
with $\nu_{zj}(\theta_0) = \operatorname{var}(O_{zij}(\theta_0)\mid Z_i=z)$, and
$\gamma_j(\theta_0) =
[\mathbb{E}\{\operatorname{var}(X_i\mid Z_i)\}]^{-1}
\sum_z p_z \operatorname{cov}(X_i, O_{zij}(\theta_0)\mid Z_i = z)$
the population within-stratum regression vector (Lemma~S1(b)
of \citealt{ye2024covariate}). Since $\sigma^2_{CSL} \le \sigma^2_{SL}$
(the variance reduction term is positive semidefinite),
$\hat\theta_{CSL}$ is at least as efficient as $\hat\theta_{SL}$.

\textbf{Variance estimator.} A consistent estimator of the asymptotic
variance in~\eqref{eq:AN_strat} is
\begin{equation}\label{eq:VarEq_strat}
\frac{\hat\sigma^2_{CSL}(\hat\theta_{CSL})}
{\{\hat\sigma^2_{SL}(\hat\theta_{CSL})\}^2},
\end{equation}
where
\[
\hat\sigma^2_{SL}(\hat\theta_{CSL})
= \frac{1}{n}\sum_z\sum_{i:\,Z_i=z}
\int_0^\tau
\frac{e^{\hat\theta_{CSL}}\bar{Y}_{z1}(t)\bar{Y}_{z0}(t)}
{\{e^{\hat\theta_{CSL}}\bar{Y}_{z1}(t)+\bar{Y}_{z0}(t)\}^2}
dN_i(t)
\]
is the observed stratified Fisher information, and
\begin{equation}\label{eq:CL_Var_strat}
\hat\sigma^2_{CSL}(\hat\theta_{CSL})
= \hat\sigma^2_{SL}(\hat\theta_{CSL})
- \hat\pi(1-\hat\pi)
\bigl(\hat\gamma_1(\hat\theta_{SL})
+\hat\gamma_0(\hat\theta_{SL})\bigr)^\top
\widehat{\mathbb{E}\{\operatorname{var}(X_i\mid Z_i)\}}
\bigl(\hat\gamma_1(\hat\theta_{SL})
+\hat\gamma_0(\hat\theta_{SL})\bigr).
\end{equation}
Consistency of~\eqref{eq:VarEq_strat} follows from
$\hat\gamma_j(\hat\theta_{SL}) \to \gamma_j(\theta_0)$ in probability
(Lemma~S1(b) of \citealt{ye2024covariate}) and the continuous mapping
theorem. An asymptotic $(1-\alpha)$-level confidence interval for
$\theta_0$ is
$\hat\theta_{CSL} \pm z_{\alpha/2}\cdot
\widehat{\operatorname{Var}}(\hat\theta_{CSL})^{1/2}$.

\subsubsection{Variance Reduction with a Scalar Prognostic Score}\label{stratified-variance-reduction-with-a-scalar-prognostic-score}

Suppose $X_i = \eta(X_i)\in\mathbb{R}$ is the pre-trained scalar
prognostic score.  With a scalar covariate, the pooled within-stratum
regression vector reduces to the scalar
\[
  \gamma_j
  = \frac{\sum_z p_z\operatorname{cov}(\eta(X_i),\,O_{zij}(\theta_0)\mid Z_i = z)}
         {\mathbb{E}\{\operatorname{var}(\eta(X_i)\mid Z_i)\}},
\]
where the numerator is the pooled within-stratum covariance.
The variance reduction term in~\eqref{eq:sigCSL} then becomes
\begin{equation}\label{eq:VR_scalar_raw}
  \pi(1-\pi)(\gamma_1+\gamma_0)^2
  \mathbb{E}\{\operatorname{var}(\eta(X_i)\mid Z_i)\}
  = \frac{\pi(1-\pi)\,[\sum_z p_z \operatorname{cov}(\eta(X_i),O_{zi1}\mid Z_i = z)
                       +\sum_z p_z \operatorname{cov}(\eta(X_i),O_{zi0}\mid Z_i = z)]^2}
         {\mathbb{E}\{\operatorname{var}(\eta(X_i)\mid Z_i)\}}.
\end{equation}

We now define the pooled within-stratum correlation for arm $j$ as
\begin{equation}\label{eq:rho_strat_def}
  \rho_{\mathrm{strat},j}
  \;=\;
  \frac{\sum_z p_z \operatorname{cov}(\eta(X_i),\,O_{zij}(\theta_0)\mid Z_i = z)}
       {\sqrt{\mathbb{E}\{\operatorname{var}(\eta(X_i)\mid Z_i)\}}
        \cdot\sqrt{\bar\nu_j(\theta_0)}},
\end{equation}
where $\bar\nu_j(\theta_0) = \sum_z p_z\nu_{zj}(\theta_0)$ and
$\nu_{zj}(\theta_0)=\operatorname{var}(O_{zij}(\theta_0)\mid Z_i=z)$.
This is the natural within-stratum analogue of the marginal correlation
$\rho_j = \operatorname{cov}(\eta(X_i),O_{ij})/\sqrt{\operatorname{var}(\eta)\cdot\nu_j}$
used in Section~3.1.
Substituting into~\eqref{eq:VR_scalar_raw}:
\begin{equation}\label{eq:VR_scalar_sub}
  \sigma^2_{SL} - \sigma^2_{CSL}
  = \pi(1-\pi)\!\left(
      \rho_{\mathrm{strat},1}\sqrt{\textstyle\bar\nu_1(\theta_0)}
     +\rho_{\mathrm{strat},0}\sqrt{\textstyle\bar\nu_0(\theta_0)}
    \right)^{\!2},
\end{equation}
and the exact stratified variance reduction ratio is
\begin{equation}\label{eq:VRR_strat}
  \frac{\sigma^{2}_{CSL}(\theta_0)}{\sigma^{2}_{SL}(\theta_0)}
  = 1 -
    \frac{\pi(1-\pi)\!\left(
        \rho_{\mathrm{strat},1}\sqrt{\bar\nu_1(\theta_0)}
       +\rho_{\mathrm{strat},0}\sqrt{\bar\nu_0(\theta_0)}
      \right)^{\!2}}
    {\displaystyle\sum_z p_z\bigl\{
       \pi\nu_{z1}(\theta_0)+(1-\pi)\nu_{z0}(\theta_0)
     \bigr\}}.
\end{equation}

\textbf{Variance reduction under the null and local alternatives.}
As in Section~3.1, we evaluate~\eqref{eq:VRR_strat} under
$H_0\colon\theta_0=0$, where the pseudo-outcomes have a known
limiting structure. Under $H_0$, the weight
$w_j(t;0) \to c_j(0)$ as $n\to\infty$ within each stratum, with the
same arm-specific constants $c_1(0) = 1-\pi$ and $c_0(0) = \pi$ as in
Section~3.1 (they do not depend on the stratum $z$, since under
Condition~A1, $\mu_z(t) = \Pr(I_i=1\mid Y_i(t)=1, Z_i=z) \to \pi$
uniformly in $z$). Therefore, within each stratum $z$:
\[
  O_{zi1}(0) \to (1-\pi)\,M_{zi1}(\tau),
  \qquad
  O_{zi0}(0) \to \pi\,M_{zi0}(\tau),
\]
where $M_{zij}(\tau) = N_{ij}(\tau) - \int_0^\tau Y_{ij}(s)\,d\hat\Lambda_{z}(s)$ 
is the arm-$j$ martingale residual for patients
in stratum $z$. Consequently, letting
$\nu_{z,M} = \operatorname{var}(M_{zij}(\tau)\mid Z_i=z)$ denote the
common within-stratum martingale variance (equal across arms under
$H_0$):
\begin{equation}\label{eq:nu_strat_H0}
  \nu_{z1}(0) \to (1-\pi)^2\,\nu_{z,M},
  \qquad
  \nu_{z0}(0) \to \pi^2\,\nu_{z,M}.
\end{equation}

The pooled within-stratum correlations satisfy
\begin{equation}\label{eq:rho_strat_H0}
  \rho_{\mathrm{strat},j}(0)
  = \operatorname{Corr}\!\bigl(\eta(X_i),\,c_j(0)\,M_{zij}(\tau) \bigr)
  = \operatorname{Corr}\!\bigl(\eta(X_i),\,M_{zij}(\tau) \bigr)
  =: \rho_{\mathrm{strat}},
\end{equation}
where the $c_j(0)$ factor diminishes because correlations are
invariant to positive scaling. The definition of correlation is
analog to equation \ref{eq:rho_strat_def}.
This quantity does not depend on arm $j$ for two reasons: (i)~under
$H_0$, both arms share the same conditional survival distribution
within each stratum, so
$\operatorname{cov}(\eta(X_i), M_{zij}(\tau)\mid Z_i=z)$ and
$\operatorname{var}(M_{zij}(\tau)\mid Z_i=z)$ are identical for
$j=0$ and $j=1$; and (ii)~$X_i\perp I_i\mid Z_i$ under
Condition~A1, so the within-stratum covariance between $\eta(X_i)$
and $M_{zij}(\tau)$ does not depend on which arm a patient is assigned
to.  The scalar $\rho_{\mathrm{strat}}$ is therefore the pooled
within-stratum correlation between the prognostic score and the
martingale residual.

Substituting~\eqref{eq:nu_strat_H0} and~\eqref{eq:rho_strat_H0}
into~\eqref{eq:VRR_strat}, with $\bar\nu_{M}=\sum_z p_z\nu_{z,M}$, 
we have
\begin{equation}\label{eq:VRR_strat_H0}
  \frac{\sigma^{2}_{CSL}(\theta_0)}{\sigma^{2}_{SL}(\theta_0)}
  = 1 - \frac{\pi(1-\pi)\,\rho_{\mathrm{strat}}^2\,
    \Bigl(
      (1-\pi)\sqrt{\bar\nu_M} + \pi\sqrt{\bar\nu_M}
    \Bigr)^2}{\sum_z p_z\bigl\{\pi(1-\pi)^2\nu_{z,M}+(1-\pi)\pi^2\nu_{z,M}\bigr\}}
  = 1 - \rho_{\mathrm{strat}}^2.
\end{equation}

Under the local alternative $\theta_0 = cn^{-1/2}$, Lemma~S4 of
\citet{ye2024covariate} applies stratum by stratum:
$\nu_{zj}(\theta_0)\to\nu_{zj}(0)$ and
$\rho_{\mathrm{strat},j}(\theta_0)\to\rho_{\mathrm{strat},j}(0)$,
so formula~\eqref{eq:VRR_strat_H0} holds asymptotically under local
alternatives as well. We therefore conclude that under $H_0$ or local
alternatives:
\begin{equation}\label{eq:VRR_strat_main}
  \frac{\operatorname{Var}(\hat\theta_{CSL})}
       {\operatorname{Var}(\hat\theta_{SL})}
  = \frac{\sigma^{2}_{CSL}(\theta_0)}{\sigma^{2}_{SL}(\theta_0)}
  \approx 1 - \rho_{\mathrm{strat}}^2,
\end{equation}
the direct stratified analogue of equation~\eqref{eq:VRR_main}.

\subsubsection{Event Number Reduction under Stratified Randomization}\label{event-number-reduction-under-stratified-randomization}

By the same argument as Section~3.2, with the non-centrality of
${U}_{CSL}(0)$ scaling as $\sqrt{d}/\sigma_{CSL}$ and that of
${U}_{SL}$ as $\sqrt{d}/\sigma_{SL}(0)$, the required event counts
satisfy
\begin{equation}\label{eq:event_reduction_strat}
  \frac{d_{\text{CSL}}}{d_{\text{SL}}}
  = \frac{\sigma^{2}_{CSL}(0)}{\sigma^{2}_{SL}(0)}
  = 1 - \rho^{2}_{\text{strat}},
\end{equation} and the number of events saved relative to the unadjusted
stratified test is 
\[
  \text{Events saved} = \rho^2_{\text{strat}} \times d_{\text{SL}}.
\] 
This is the direct stratified analogue of the event savings formula
in Section~3.2. The gain $\rho^2_{\text{strat}}$ is in general no larger
than the corresponding $\rho^2$ under simple randomization with the same
score, since stratification absorbs some of the variation that the
prognostic score would otherwise explain. In the special case where
$Z_i$ and $\eta(X_i)$ are independent, $\rho_{\text{strat}} = \rho$ and
the two formulas coincide.

\newpage
\begin{landscape}

\subsection{Additional Simulation Results}\label{additional-simulation-results}

Table~\ref{tb:SimuII} reports additional simulation results comparing
four analysis strategies: (i)~adjusting for the martingale-residual
prognostic score alone ($\sim\eta^{(\hat{M})}$); (ii)~adjusting for
the martingale-residual score plus all observed baseline covariates
($\sim X_1+X_2+X_3+\eta^{(\hat{M})}$); (iii)~adjusting for the
survival-probability prognostic score alone ($\sim\eta^{(\hat{S})}$);
and (iv)~adjusting for the survival-probability score plus all baseline
covariates ($\sim X_1+X_2+X_3+\eta^{(\hat{S})}$). The simulation
setting is as described in Section~4.1. Results for strategy~(i) are
identical to those in Table~\ref{tb:SimuI} and are repeated here for
ease of comparison.

The main findings are as follows. When the prognostic score already
captures the major external prognostic information (Cases~I,~VI and~VII),
adding observed baseline covariates provides little additional gain.
However, in cases where the external dataset is only partially
informative or the key covariate is missing from the historical data
(Cases~II--V), adding baseline covariates alongside the prognostic score
leads to meaningful additional power gain, by leveraging the internal
prognostic information from the concurrent trial. This supports the
practical recommendation to include the prognostic score as a supplement
to, rather than a replacement for, observed baseline covariates.

With respect to the choice of training target, predicting the
martingale residual $\hat{M}^{\text{ext}}(\tau)$ and predicting the
survival probability $\hat{S}(t)$ at the median follow-up time yield
comparable efficiency gains across all cases. The survival probability
target may be preferable in practice due to its familiarity and the
availability of well-established software for survival random forests.

\begingroup\fontsize{9}{11}\selectfont

\begin{longtable}[t]{cccccccccccccc}
\caption{\label{tab:HRinfoBorrowII}\label{tb:SimuII}Additional simulation results comparing four analysis strategies. $\sim\eta^{(\hat{M})}$: adjust for prognostic score trained on martingale residual only (same as Table~\ref{tb:SimuI}). $\sim X_1+X_2+X_3+\eta^{(\hat{M})}$: adjust for martingale-residual score plus all baseline covariates. $\sim\eta^{(\hat{S})}$: adjust for prognostic score trained on survival probability only. $\sim X_1+X_2+X_3+\eta^{(\hat{S})}$: adjust for survival-probability score plus all baseline covariates. Columns as in Table~\ref{tb:SimuI}.}\\
\toprule
\multicolumn{2}{c}{ } & \multicolumn{3}{c}{$\sim \eta^{(\hat{M})}$} & \multicolumn{3}{c}{$\sim X_1+X_2+X_3+\eta^{(\hat{M})}$} & \multicolumn{3}{c}{$\sim \eta^{(\hat{S})}$} & \multicolumn{3}{c}{$\sim X_1+X_2+X_3+\eta^{(\hat{S})}$} \\
\cmidrule(l{3pt}r{3pt}){3-5} \cmidrule(l{3pt}r{3pt}){6-8} \cmidrule(l{3pt}r{3pt}){9-11} \cmidrule(l{3pt}r{3pt}){12-14}
 & $n$ & Bias & MSE & Pr.\ Rej.\ H0 & Bias & MSE & Pr.\ Rej.\ H0 & Bias & MSE & Pr.\ Rej.\ H0 & Bias & MSE & Pr.\ Rej.\ H0\\
\midrule
\addlinespace[0.3em]
\multicolumn{14}{l}{\textbf{Case I:}}\\
\hspace{1em} & 200 & 0.001 & 0.123 & 0.046 & 0.001 & 0.122 & 0.051 & 0.001 & 0.123 & 0.048 & 0.001 & 0.121 & 0.051\\
\cmidrule{2-14}\nopagebreak
\hspace{1em}\multirow{-2}{*}{\centering\arraybackslash Null} & 400 & 0.000 & 0.087 & 0.051 & 0.000 & 0.086 & 0.054 & 0.001 & 0.087 & 0.051 & 0.000 & 0.085 & 0.053\\
\cmidrule{1-14}\pagebreak[0]
\hspace{1em} & 200 & 0.001 & 0.131 & 0.645 & 0.001 & 0.129 & 0.654 & 0.001 & 0.130 & 0.650 & 0.001 & 0.128 & 0.660\\
\cmidrule{2-14}\nopagebreak
\hspace{1em}\multirow{-2}{*}{\centering\arraybackslash Efficacy} & 400 & 0.000 & 0.092 & 0.914 & 0.000 & 0.091 & 0.916 & 0.000 & 0.092 & 0.911 & 0.000 & 0.091 & 0.915\\
\cmidrule{1-14}\pagebreak[0]
\addlinespace[0.3em]
\multicolumn{14}{l}{\textbf{Case II:}}\\
\hspace{1em} & 200 & 0.000 & 0.156 & 0.052 & 0.001 & 0.146 & 0.056 & 0.000 & 0.163 & 0.053 & 0.001 & 0.147 & 0.056\\
\cmidrule{2-14}\nopagebreak
\hspace{1em}\multirow{-2}{*}{\centering\arraybackslash Null} & 400 & 0.000 & 0.110 & 0.050 & 0.000 & 0.104 & 0.052 & 0.000 & 0.115 & 0.048 & 0.000 & 0.104 & 0.051\\
\cmidrule{1-14}\pagebreak[0]
\hspace{1em} & 200 & 0.000 & 0.161 & 0.474 & 0.000 & 0.152 & 0.522 & 0.001 & 0.168 & 0.449 & 0.000 & 0.153 & 0.521\\
\cmidrule{2-14}\nopagebreak
\hspace{1em}\multirow{-2}{*}{\centering\arraybackslash Efficacy} & 400 & 0.000 & 0.114 & 0.759 & 0.001 & 0.108 & 0.807 & 0.000 & 0.119 & 0.723 & 0.001 & 0.108 & 0.803\\
\cmidrule{1-14}\pagebreak[0]
\addlinespace[0.3em]
\multicolumn{14}{l}{\textbf{Case III:}}\\
\hspace{1em} & 200 & 0.000 & 0.167 & 0.052 & 0.001 & 0.148 & 0.056 & 0.000 & 0.167 & 0.054 & 0.001 & 0.148 & 0.055\\
\cmidrule{2-14}\nopagebreak
\hspace{1em}\multirow{-2}{*}{\centering\arraybackslash Null} & 400 & 0.000 & 0.118 & 0.049 & 0.001 & 0.105 & 0.051 & 0.000 & 0.118 & 0.048 & 0.001 & 0.105 & 0.051\\
\cmidrule{1-14}\pagebreak[0]
\hspace{1em} & 200 & 0.001 & 0.172 & 0.431 & 0.000 & 0.153 & 0.518 & 0.001 & 0.172 & 0.432 & 0.000 & 0.153 & 0.516\\
\cmidrule{2-14}\nopagebreak
\hspace{1em}\multirow{-2}{*}{\centering\arraybackslash Efficacy} & 400 & 0.001 & 0.121 & 0.712 & 0.002 & 0.109 & 0.801 & 0.000 & 0.122 & 0.708 & 0.001 & 0.109 & 0.801\\
\cmidrule{1-14}\pagebreak[0]
\addlinespace[0.3em]
\multicolumn{14}{l}{\textbf{Case IV:}}\\
\hspace{1em} & 200 & 0.000 & 0.166 & 0.053 & 0.001 & 0.147 & 0.056 & 0.000 & 0.167 & 0.055 & 0.001 & 0.146 & 0.057\\
\cmidrule{2-14}\nopagebreak
\hspace{1em}\multirow{-2}{*}{\centering\arraybackslash Null} & 400 & 0.000 & 0.117 & 0.050 & 0.001 & 0.104 & 0.051 & 0.000 & 0.118 & 0.049 & 0.001 & 0.104 & 0.051\\
\cmidrule{1-14}\pagebreak[0]
\hspace{1em} & 200 & 0.001 & 0.171 & 0.437 & 0.000 & 0.153 & 0.519 & 0.001 & 0.172 & 0.435 & 0.001 & 0.152 & 0.521\\
\cmidrule{2-14}\nopagebreak
\hspace{1em}\multirow{-2}{*}{\centering\arraybackslash Efficacy} & 400 & 0.000 & 0.121 & 0.714 & 0.001 & 0.108 & 0.803 & 0.000 & 0.121 & 0.710 & 0.001 & 0.108 & 0.805\\
\cmidrule{1-14}\pagebreak[0]
\addlinespace[0.3em]
\multicolumn{14}{l}{\textbf{Case V:}}\\
\hspace{1em} & 200 & 0.001 & 0.147 & 0.051 & 0.001 & 0.125 & 0.056 & 0.001 & 0.145 & 0.049 & 0.001 & 0.130 & 0.057\\
\cmidrule{2-14}\nopagebreak
\hspace{1em}\multirow{-2}{*}{\centering\arraybackslash Null} & 400 & 0.001 & 0.104 & 0.047 & 0.000 & 0.088 & 0.052 & 0.000 & 0.102 & 0.050 & 0.000 & 0.092 & 0.053\\
\cmidrule{1-14}\pagebreak[0]
\hspace{1em} & 200 & 0.001 & 0.153 & 0.515 & 0.001 & 0.132 & 0.636 & 0.001 & 0.151 & 0.527 & 0.000 & 0.136 & 0.612\\
\cmidrule{2-14}\nopagebreak
\hspace{1em}\multirow{-2}{*}{\centering\arraybackslash Efficacy} & 400 & 0.000 & 0.108 & 0.807 & 0.001 & 0.093 & 0.903 & 0.000 & 0.106 & 0.815 & 0.001 & 0.097 & 0.882\\
\cmidrule{1-14}\pagebreak[0]
\addlinespace[0.3em]
\multicolumn{14}{l}{\textbf{Case VI:}}\\
\hspace{1em} & 200 & 0.003 & 0.094 & 0.048 & 0.003 & 0.092 & 0.054 & 0.002 & 0.112 & 0.053 & 0.002 & 0.109 & 0.057\\
\cmidrule{2-14}\nopagebreak
\hspace{1em}\multirow{-2}{*}{\centering\arraybackslash Null} & 400 & 0.000 & 0.067 & 0.051 & 0.000 & 0.065 & 0.053 & 0.000 & 0.079 & 0.050 & 0.000 & 0.078 & 0.054\\
\cmidrule{1-14}\pagebreak[0]
\hspace{1em} & 200 & 0.002 & 0.101 & 0.890 & 0.002 & 0.099 & 0.903 & 0.002 & 0.119 & 0.777 & 0.001 & 0.116 & 0.793\\
\cmidrule{2-14}\nopagebreak
\hspace{1em}\multirow{-2}{*}{\centering\arraybackslash Efficacy} & 400 & 0.001 & 0.071 & 0.994 & 0.001 & 0.070 & 0.995 & 0.000 & 0.084 & 0.966 & 0.001 & 0.082 & 0.971\\
\cmidrule{1-14}\pagebreak[0]
\addlinespace[0.3em]
\multicolumn{14}{l}{\textbf{Case VII:}}\\
\hspace{1em} & 200 & 0.001 & 0.123 & 0.052 & 0.001 & 0.121 & 0.055 & 0.001 & 0.123 & 0.055 & 0.001 & 0.121 & 0.055\\
\cmidrule{2-14}\nopagebreak
\hspace{1em}\multirow{-2}{*}{\centering\arraybackslash Null} & 400 & 0.000 & 0.086 & 0.049 & 0.000 & 0.086 & 0.051 & 0.001 & 0.087 & 0.049 & 0.001 & 0.085 & 0.051\\
\cmidrule{1-14}\pagebreak[0]
\hspace{1em} & 200 & 0.001 & 0.131 & 0.711 & 0.001 & 0.129 & 0.716 & 0.001 & 0.131 & 0.706 & 0.001 & 0.129 & 0.719\\
\cmidrule{2-14}\nopagebreak
\hspace{1em}\multirow{-2}{*}{\centering\arraybackslash Efficacy} & 400 & 0.000 & 0.092 & 0.944 & 0.000 & 0.091 & 0.944 & 0.000 & 0.092 & 0.941 & 0.000 & 0.091 & 0.947\\
\bottomrule
\end{longtable}
\endgroup{}

\end{landscape}

\bibliographystyle{unsrtnat}
\bibliography{ref} 

\end{document}